\documentclass[11pt,a4paper]{article}                           

\usepackage[bookmarksopen, bookmarksnumbered, bookmarksopenlevel=2]{hyperref}
\usepackage{tikz}
\usepackage{tikz-3dplot}
\usetikzlibrary{calc}
\usetikzlibrary{decorations} %
\usepackage[UKenglish]{babel}
\usepackage[toc,page]{appendix}
\usepackage{amsmath}
\usepackage{amssymb}
\usepackage{graphicx}
\usepackage{hhline}
\usepackage[bf]{caption}
\usepackage{cite}
\usepackage[vcentermath]{youngtab}
\usepackage{geometry}
\DeclareSymbolFont{bbold}{U}{bbold}{m}{n}
\DeclareSymbolFontAlphabet{\mathbbold}{bbold}

 \def\be{\begin{equation}}
\def\ee{\end{equation}}
\def\bseq{\begin{subequations}}
\def\eseq{\end{subequations}}

\def\bea{\begin{eqnarray}}
\def\eea{\end{eqnarray}}

\def\bseq{\begin{subequations}}
\def\eseq{\end{subequations}}

\usepackage{todonotes}

\hypersetup{
    pdftitle={M2-M5 in f-theory},
    pdfauthor={Luca Martucci, Timo Weigand},
    pdfsubject={F-theory compactifications}
}
\numberwithin{equation}{section}
\numberwithin{table}{section}

\newcommand{\tw}{{\rm w}}

\newcommand{\executeiffilenewer}[3]{%
 \ifnum\pdfstrcmp{\pdffilemoddate{#1}}%
 {\pdffilemoddate{#2}}>0%
 {\immediate\write18{#3}}\fi%
}

\def\X {X}
\def\Y {Y}
\def\B  {B}

\def\d {{\rm d}}

\def\calc         {{\cal C}}
\def\cald         {{\cal D}}

\def\calf         {{\cal F}}
\def\calg         {{\cal G}}

\def\call         {{\cal L}}

\def\cals         {{\cal S}}

\def\tw   {{\rm w}}

\def\del          {\partial}

\def\ii           {{\rm i}}

\def\sqr#1#2{{\vcenter{\vbox{\hrule height.#2pt
 \hbox{\vrule width.#2pt height#1pt \kern#1pt \vrule width.#2pt}\hrule
 height.#2pt}}}}

\def\tb{\rm b}

\def\d{\text{d}}

\def\slashchar#1{\setbox0=\hbox{$#1$}           
\dimen0=\wd0                                 
\setbox1=\hbox{/} \dimen1=\wd1               
\ifdim\dimen0>\dimen1                        
\rlap{\hbox to \dimen0{\hfil/\hfil}}      
#1                                        
\else                                        
\rlap{\hbox to \dimen1{\hfil$#1$\hfil}}   
/                                         
\fi}

\begin{document}

\baselineskip=14pt
\parskip 5pt plus 1pt

\vspace*{-1.5cm}
\begin{flushright}    
  {\tiny DFPD-2015/TH/18

  }
\end{flushright}

\vspace{2cm}
\begin{center}        
  {\LARGE Hidden Selection Rules, M5-instantons and Fluxes in F-theory}
\end{center}

\vspace{0.75cm}
\begin{center}        
Luca Martucci$^1$ and  Timo Weigand$^2$
\end{center}

\vspace{0.15cm}
\begin{center}        
\emph{ 
$^1$ Dipartimento di Fisica e Astronomia `Galileo Galilei', Universit\`a di Padova, \\
\& I.N.F.N.\ Sezione di Padova, via Marzolo 8, I-35131 Padova, Italy\\
\smallskip
$^2$Institut f\"ur Theoretische Physik, Ruprecht-Karls-Universit\"at, \\
             Philosophenweg 19, 69120
             Heidelberg, Germany
             }
\end{center}

\vspace{2cm}


\begin{abstract}
\noindent  We introduce a new approach  to investigate the selection rules governing the contributions of fluxed M5-instantons to the F-theory four-dimensional effective action, with emphasis on the generation of charged matter F-terms.
The structure of such couplings is unraveled by exploiting the perturbative and non-perturbative homological relations, introduced in our companion paper  \cite{Martucci:2015dxa}, which encode the interplay between the self-dual 3-form flux on the M5-brane, the background 4-form flux and certain
fibral curves. The latter are wrapped by time-like M2-branes representing matter insertions in the instanton path integral. 
In particular, we clarify how  fluxed M5-instantons detect the presence of geometrically massive $U(1)$s which are responsible for `hidden' selection rules.  
We discuss how for non-generic embeddings the M5-instanton can probe `locally massless' $U(1)$ symmetries if the rank of its Mordell-Weil group is enhanced compared to that of the bulk.
As a phenomenological off-spring we propose a new type of non-perturbative corrections to Yukawa couplings which may change the rank of the Yukawa matrix. 
Along the way, we also gain new insights into the structure of massive $U(1)$ gauge fluxes in the stable degeneration limit.

\end{abstract}

\thispagestyle{empty}
\clearpage
\setcounter{page}{1}


\newpage

\tableofcontents

\section{Introduction}

  Instanton effects play a crucial role in string compactifications. In Type IIB/F-theory Calabi-Yau compactifications, D3/M5-brane instantons are the only source of a superpotential for the K\"ahler moduli \cite{Witten:1996bn} and thus represent vital ingredients for moduli stabilisation. In addition they can generate perturbatively forbidden matter couplings in the effective action \cite{Blumenhagen:2006xt,Ibanez:2006da,Florea:2006si,Haack:2006cy} whose rich phenomenology has been explored extensively in the weak coupling regime  (see e.g. \cite{Blumenhagen:2009qh} and references therein).
 It is the goal of this paper to advance our theoretical understanding of corresponding effects in the much larger class of general F-theory compactifications which do not necessarily admit a weak coupling limit.
While many aspects of  such instanton effects in F-theory have been investigated  recently  \cite{Heckman:2008es,Marsano:2008py,Cvetic:2009ah,Blumenhagen:2010ja,Cvetic:2010rq,Donagi:2010pd,Cvetic:2010ky,Blumenhagen:2010ed,Marsano:2011nn,Cvetic:2011gp,Grimm:2011sk,Bianchi:2011qh,Kerstan:2012cy,Cvetic:2012ts,Bianchi:2012kt,Ko:2013dka,Martucci:2014ema}, it is fair to say that an understanding at a comparable level as in Type IIB orientifolds is still lacking.
This is partly because the physics of M5-instantons is complicated by the appearance of a self-dual 2-form in its worldvolume \cite{Witten:1996hc,Pasti:1997gx}. Relatedly, the structure of the charged fermionic zero modes responsible for the generation of charged matter couplings is much more obscure than at weak coupling. Luckily the information relevant for the computation of such operators is encoded already in the classical Bianchi identity of the 3-form field strength $T_3$ associated with the chiral 2-form on the M5-brane.

This Bianchi identity describes the interplay between the instanton 3-form flux $T_3$, the gauge flux $G_4$ in the F-theory bulk and further time-like M2-branes that can end on the M5-brane instanton. The latter wrap a combination $C$ of vanishing cycles in the torus fibre of the F/M-theory Calabi-Yau 4-fold. Such wrapped M2-branes describe massless charged matter fields in the F-theory effective action \cite{Witten:1996qb,Katz:1996xe,Intriligator:1997pq,LieF}. The Bianchi identity then has the structure
\be\label{BI0}
\d T_3 = G_4|_{\rm M5} - \delta_{4} (C),
\ee
where $\delta_{4} (C)$ is a delta-like 4-form supported on $C$ which represents a localised source  for the $T_3$ flux. 
At the crudest cohomological level, the Bianchi identity (\ref{BI0}) implies that the difference of $G_4|_{\rm M5}-{\rm PD}[C]$ must be cohomologically trivial, with ${\rm PD}[C]$ denoting the Poincar\'e dual of $C$ on the M5-brane. 
In particular, if $ G_4|_{\rm M5}$ is cohomologically non-trivial, then $C$ must be homologically non-trivial, too, and the M5-brane instanton contributes to couplings involving the states associated with the M2-branes wrapping $C$.
While the role of this  Bianchi identity for the generation of instanton couplings has been analysed before \cite{Donagi:2010pd,Marsano:2011nn,Kerstan:2012cy}, the novelty of our approach is to solve it not just at the (co)homological level, but in terms of a refined version of homology introduced in our companion paper \cite{Martucci:2015dxa}. This allows us to describe qualitatively new M5-instanton couplings and to understand in particular the generation of instanton induced couplings violating massive $U(1)$ symmetries \cite{Grimm:2010ez,Grimm:2011tb,Grimm:2011dj,Braun:2014nva}. 

Our starting point is the following observation. Recall that an M5-brane instanton in F/M-theory uplifting a D3-instanton in the dual Type IIB theory must wrap a {\em vertical} divisor $D$  \cite{Witten:1996bn} of the F-theory genus-one fibration, which is itself a genus-one fibered K\"ahler space. The pullback of the 4-form flux $G_4$ to the instanton is dual to a curve $C_G$ on the M5-brane world-volume. Hence, at the homological level, the Bianchi identity (\ref{BI0}) implies that the difference of $C_G - C$ must be homologically trivial on $D$.
However, only if this cancellation occurs also in the  homology of the instanton \emph{fibre} \cite{Martucci:2015dxa}  can the instanton 3-form flux vanish identically. Otherwise, the M5-path integral is consistent only if a suitable supersymmetric $T_3$ flux exists. Such flux may be thought of as  Poincar\'e dual to a 3-chain $\Gamma\subset D$ ending  on the difference of cycles $C_G-C$ and the resulting fluxed M5-brane can then be viewed as a BPS  bound state of an M5-instanton with an M2-instanton wrapping the 3-chain $\Gamma$. Interestingly, while M2-brane instantons on 3-chains cannot contribute to F-terms in F-theory because they are non-BPS on Calabi-Yau 4-folds, fluxed M5-instantons can be BPS.
This way we find the M5-instanton analogue of a number of F-term effects which had previously only been understood in the language of fluxed D3-brane instantons in Type IIB orientifolds \cite{Grimm:2011tb}. 
For convenience of the reader, we summarise the most salient points concerning such fluxed D3-instantons in appendix \ref{chargedD3inst}.

We begin the core of this article in section \ref{sec:M5action} by defining the general type of Euclidean fluxed M5-configurations we are interested in. This includes a systematic analysis of the supersymmetry condition for the 3-form flux.
In section \ref{sec_pertvsnonpert} we explain the notion of perturbative versus non-perturbative homological relations on the instanton divisor, thereby adopting the concepts introduced in \cite{Martucci:2015dxa}.  Two curves that are homologous  in $D$ are said to be equivalent in the {\em perturbative} homology of $D$ if they can be connected by a 3-chain $\Gamma$ whose volume vanishes in the F-theory limit.
Otherwise, they are said to be {\em non-perturbatively} equivalent. 
These concepts can be applied both to the fibral curve $C$ wrapped by time-like M2-branes and to the 2-cycle $C_G$ dual to the pullback  of the background flux. We thereby make natural contact with the description of such fluxes as algebraic 4-cycles \cite{Braun:2011zm}, more precisely as rational equivalence classes thereof \cite{Bies:2014sra}.
 
 Our first application of these concepts, described in section \ref{sec:G4trivial}, is to M5-instantons in backgrounds with vanishing gauge flux $G_4$. Clearly, in this case, $C$ must be homologically trivial on $D$. However, as we shall see, an M5-instanton with non-trivial $T_3$ flux can still generate operators in the effective action with net charge under a geometrically massive $U(1)$ symmetry in the sense of \cite{Grimm:2010ez,Grimm:2011tb,Grimm:2011dj,Braun:2014nva}.  The states appearing in these operators are due to M2-branes wrapping fibral curves of the instanton divisor which are indeed trivial in the non-perturbative homology, but {\em not} in the perturbative one. An example of such a fibral curve arises in the massive $U(1)$ model of \cite{Braun:2014nva}. 
Another interesting   application of potential phenomenological importance is the generation of non-perturbative corrections to perturbatively allowed matter Yukawa couplings. Non-perturbative corrections can change the rank of the Yukawa matrix and account for the observed family  hierarchies \cite{Marchesano:2009rz,Aparicio:2011jx,Font:2012wq,Font:2013ida,Marchesano:2015dfa}. 
We here present a qualitatively new such correction relying on the described mechanism involving $T_3$ flux on the instanton.

Even for vanishing $T_3$ our approach to solving the Bianchi identity gives new insights. Such configurations are possible if one cancels the contribution $C_G$ due to background gauge flux in the perturbative homology against M2-branes wrapping a fibral curve. 
It is well-known \cite{Donagi:2010pd,Marsano:2011nn,Kerstan:2012cy} that a homologically non-trivial flux induces a charge for a massless $U(1)$ of the instanton analogous to corresponding effects in Type II  \cite{Blumenhagen:2006xt,Ibanez:2006da,Florea:2006si,Haack:2006cy}. This alone, however, does not completely determine the form of the actual operator induced in the effective action.
In section \ref{sec:T3=0} we show for an $SU(5)$ F-theory model how to identify this operator in more precise terms. Importantly, this analysis is applicable also to situations where the flux-induced instanton charge with respect to a massless $U(1)$ vanishes, but an operator involving localised matter is nonetheless induced for $T_3=0$. 

Finally, in section \ref{sec_thirdcase}, we exemplify an interesting situation in which the background gauge flux and the instanton flux compensate each other such that a contribution to the superpotential without charged matter insertions is possible. This realises an effect proposed in the Type IIB language in \cite{Grimm:2011tb} for models with a geometrically massive $U(1)$. Necessarily, the curve $C_G$ dual to the background flux must be homologically trivial in a non-perturbative sense. Since the global description of massive $U(1)$ gauge fluxes is still largely elusive as of this writing (see however \cite{Krause:2012yh}) we address this question in the stable degeneration limit  \cite{Clingher:2012rg,Donagi:2012ts} of the massive $U(1)$ model of \cite{Braun:2014nva}. Along the way we gain a deeper understanding of the structure of massive $U(1)$ fluxes, in particular of the D5-tadpole cancellation condition as a requirement imposed by the Meyer-Vietoris exact sequence in the stable degeneration. 

In all the applications described up to this points we have assumed that the instanton wraps a sufficiently generic, smooth vertical divisor. Interesting new effects occur in non-generic situations in which for instance the rank of the Mordell-Weil group along the vertical divisor increases. The existence of extra sections along the instanton implies that the instanton probes a `locally massless $U(1)$' which is Higgsed or St\"uckelberg massive globally away from the instanton. We exemplify this in section \ref{sec:gaugenh} by discussing such a local $U(1)$ enhancement along an instanton in a Tate model and in the bisection fibration studied in \cite{Braun:2014oya,Morrison:2014era,Anderson:2014yva,Klevers:2014bqa,Garcia-Etxebarria:2014qua,Mayrhofer:2014haa,Mayrhofer:2014laa}. In the latter case the global  $\mathbb Z_2$ symmetry of the F-theory effective action is enhanced to a local $U(1)$ because the VEV of the Higgs field responsible for its breaking restricts to zero along the instanton. 
This gives rise to new selection rules for the instanton contribution to the path integral.
Our conclusions are summarised in section \ref{sec_Conclusions}.

\section{M5-brane action}
\label{sec:M5action}

In this paper we study  D3/M5-instantons  on general F-theory compactifications to four dimensions, either admitting a  weak coupling limit  or not.  In the former case, the available perturbative IIB description can significantly help in understanding and clarifying some crucial physical aspects that persist in genuinely non-perturbative F-theory vacua. 
We therefore review, in appendix \ref{chargedD3inst}, some of the most important aspects of the computation of D3-brane instanton effects in Type IIB orientifold compactifications. Such instantons wrap holomorphic divisors $E$ of the orientifold Calabi-Yau 3-fold $\X$ and can in general carry non-trivial instanton flux $F_E$. Of special interest for us is the dependence of the instanton effects on the gauge fluxes $F_a$ supported by D7-branes wrapping holomorphic divisors $D_a$  as well as on the instanton flux $F_E$. After  reviewing the distinction of the gauge flux induced and geometric St\"uckelberg mechanism rendering abelian gauge symmetries massive we recall how the D3-instanton acquires a flux-dependent $U(1)$ charge, given by equation (\ref{E3charges}). As a result the instanton contributes to operators involving matter fields charged under the abelian gauge symmetries of the form
\be \label{Phi-coup1a}
 (\Phi_{ab})^n\, e^{-S_{\rm D3}},
 \ee
where $\Phi_{ab}$ is  a matter field charged under $U(1)_a \times U(1)_b$.

Our primary goal is to understand analogous effects for more general F-theory compactifications. A type IIB D3-brane instanton is dual to an
M5-brane instanton localised at a certain external spacetime point $x_0$ and wrapping a internal {\em vertical} 6-cycle 
\bea
D = \pi^{-1} D_{\tb}
\eea
  in the elliptically\footnote{More generally, as in section \ref{sec_Z2instanton}, the F-theory background needs only to be genus-one fibered \cite{Braun:2014oya}.} fibered Calabi-Yau 4-fold $ \Y$ associated with the F-theory background \cite{Witten:1996bn}. Here $\pi:\Y\rightarrow \B$ is the projection map associated with the elliptic fibration and $D_{\tb}\subset \B$ is the effective divisor of the base $\B$ wrapped by the D3-brane. In a weak-coupling regime, the Calabi-Yau 3-fold $\X$ represents the orientifold double-cover of the base $\B$ and $D_{\tb}$ uplifts to the divisor $E\subset \X$. Many aspects of D3/M5-instantons in F-theory have recently been analysed using various techniques \cite{Heckman:2008es,Marsano:2008py,Cvetic:2009ah,Cvetic:2010ky,Blumenhagen:2010ja,Cvetic:2010rq,Donagi:2010pd,Blumenhagen:2010ed,Marsano:2011nn,Cvetic:2011gp,Grimm:2011sk,Bianchi:2011qh,Kerstan:2012cy,Cvetic:2012ts,Bianchi:2012kt,Ko:2013dka,Martucci:2014ema}.
The microscopic description of effects of the form (\ref{Phi-coup1a}) in terms of M5-instantons in F-theory is very different from the formalism available for D3-instantons at weak-coupling. One of our main goals is to work out  the connection between these two descriptions and to understand how the results obtained at weak coupling extend and generalise to more general F-theory compactifications.  

\subsection{M5-action and M2-insertions}

As bosonic action for a single (Euclidean) M5-brane instanton we take
\be \label{M5action}
\begin{aligned}
S^{(0)}_{\rm M5} &= 2 \pi\, {\rm vol}(D) +  S_{\rm k} + S_{\rm CS}, \\
S_{\rm k} &= \frac\pi2\int_D T_3\wedge *T_3, \\
S_{\rm CS} &=  -2\pi\ii  \,\int_D  \big(C_6+\frac{1}{2} T_3\wedge C_3\big), 
\end{aligned}
\ee
where the 6-form $C_6$ is the magnetic dual to the M-theory potential $C_3$ associated with the field-strength $G_4$. Let us first suppose that there are no M2-brane insertions, which shall be discussed momentarily. Then the world-volume field-strength $T_3$ must satisfy the Bianchi identity
\be\label{T3BI0}
\d T_3 = G_4|_D
\ee
so that we can locally write  
\be\label{localsplit}
T_3 = \d \beta_2 +C_3|_D
\ee
for some 2-form $\beta_2$ living on the M5-brane. 
Hence, with the help of (\ref{T3BI0}) one can verify that the combined 
 gauge transformations 
\be\label{C3gauge}
\begin{aligned}
C_3 \rightarrow C_3 + \d \Lambda_2\,,\qquad
C_6 \rightarrow C_6+\d\Lambda_5 - \frac{1}{2} \Lambda_2 \wedge G_4\,,\qquad
\beta_2 \rightarrow \beta_2 - \Lambda_2|_D\,\qquad
\end{aligned}
\ee
leave the M5-action invariant.

Notice that we are using a quadratic approximation of a more complete DBI-like action. However, since we will be interested in  supersymmetric M5-brane configurations, we expect such a quadratic approximation to capture the   information necessary for our purposes, as one can conclude from a IIB perspective \cite{Bianchi:2011qh,Bianchi:2012kt,Martucci:2014ema}.  Furthermore, it is well known that (\ref{M5action}) cannot be the end of the story since $\beta_2$ is a chiral 2-form and then must obey the imaginary self-duality condition 
\be\label{T3ISD}
*T_3 = \ii T_3\,.
\ee
 One approach, to which we will adhere, is to enforce this constraint  at the level of the partition function via holomorphic factorisation, as in \cite{Witten:1996hc}. An alternative would be to start from the formulation of  \cite{Pasti:1997gx,Bandos:1997ui}. In any case, this difficulty will not affect the following discussion. Furthermore, the action (\ref{M5action}) should be completed by its fermionic counterpart \cite{Bandos:1997ui} but we will not need such completion here.

In order to generate an operator for instance of the form (\ref{Phi-coup1a}), the M5-brane instanton must couple  to incoming M2-branes describing the physical states associated with the fields $\Phi_{ab}$. This logic has already been employed in the present context in \cite{Donagi:2010pd,Marsano:2011nn,Kerstan:2012cy}.  The incoming M2-branes move along external world-lines ending at the M5-brane external position $x_0$  and wrap internal fibral curves $C_A=\sum_i k^i_A\mathbb{P}^1_i$. Here $\mathbb{P}^1_i$ denote the resolution rational curves that are obtained by blowing-up the singular elliptic fibration $\Y$ into a smooth $\hat Y$. If we denote by $\gamma_{A,k}$, $k=1,\ldots, n_A$, the external world-lines of $n_A$ incoming M2-branes wrapping the fibral curve  $C_A$, then we can  think of the M2-branes as collectively spanning the 3-chain $P = \sum_{A,k} \gamma_{A,k}\times C_A$ ending on the M5-instanton divisor $D$.  Since $\del  \gamma_{A,k}=x_0$, we have $\del P = x_0\times C$, where  
\be
C\equiv \sum_{A} n^A C_A\subset D
\ee 
denotes the collective fibral curve in the M5-instanton on which the incoming M2-branes end.  
We focus on incoming M2-branes, but one could likewise  consider out-going M2-branes that are related to the incoming ones  by CPT.

To account for such insertions of incoming M2-branes we must modify the M5-instanton action (\ref{M5action}) into
\be\label{completeM5}
S_{\rm M5}=S^{(0)}_{\rm M5} -\pi\ii\int_C \beta_2 .
\ee
By computing the equation of motion from $S_{\rm M5}$ and using (\ref{T3ISD}) we see that the  Bianchi identity (\ref{T3BI0}) must be consistently modified into
\be\label{T3BI}
\d T_3 = G_4|_D - \delta_{4} (C).
\ee
Furthermore, one must supplement the M5-instanton action with the action of the incoming M2-branes
\be\label{M2action}
S_{\rm M2} = 2\pi {\rm vol}(P)  - 2\pi\ii\int_{P} C_3\,,
\ee
where the contribution $2\pi {\rm vol}(P)$  degenerates to the world-line action for a massless particle in the F-theory limit. 
By using the modified Bianchi identity (\ref{T3BI}), one can check that $S_{\rm M5} + S_{\rm M2}$ is invariant under the gauge transformation (\ref{C3gauge}). We stress that only the combination   $S_{\rm M5} + S_{\rm M2}$ is gauge invariant, not $S_{\rm M5}$ and  $S_{\rm M2}$ separately.

We are particularly interested in the contributions of supersymmetric M5-brane instantons to F-terms in the effective action. 
 To compute such contributions one should perform the path-integral over all bosonic and fermionic fields \cite{Witten:1996bn,Harvey:1999as,Witten:1999eg,Beasley:2005iu}  and include also the contribution of the incoming M2-branes as in  \cite{Donagi:2010pd,Marsano:2011nn,Kerstan:2012cy}. The type of F-term is partly determined by the structure of the M5-brane fermionic zero-modes, which has been investigated in  \cite{Witten:1996bn,Kallosh:2005yu,Saulina:2005ve,Kallosh:2005gs,Martucci:2005rb,Bergshoeff:2005yp,Tsimpis:2007sx,Bianchi:2011qh,Bianchi:2012kt}. On the other hand, the path-integral requires also summation over all possible gauge inequivalent configurations of the chiral form $\beta_2$. 
This path-integral splits into a classical partition function which can be evaluated as in \cite{Henningson:1999dm} together with an integral over quantum fluctuations of $\beta_2$. In particular, the latter  can lead to zeroes in the superpotential.\footnote{In IIB language, these are caused by the appearance of vectorlike fermionic zero-modes e.g. at the intersection of the D3-instanton with spacetime-filling 7-branes which cannot be absorbed in the path-integral. Such zeroes in the superpotential of D3-instantons in IIB/F-theory have recently been studied in particular in \cite{Cvetic:2012ts}. For more details on the organisation of the M5 partition function and a detailed comparison with the Type IIB formalism we refer the reader to the recent discussion \cite{Kerstan:2012cy} and references therein.}   

On the other hand, the classical piece of the M5-partition function includes  a sum over all semiclassical values of the field strength $T_3$.
As we will discuss in some detail in the following, the Bianchi identity (\ref{T3BI}) gives rise to the selection rules that constrain  the types of terms a given instanton with flux can generate in the effective action. This is already clear  by imposing (\ref{T3BI}) just at the cohomological level, i.e.\
\be\label{cohoBI}
 [G_4]_D=[C ]_D\,,
  \ee 
  where  $[G_4]_D$ and $[C]_D$ denote the cohomology classes of $G_4|_D$ and of the Poincar\'e dual of $C$, respectively, inside $D$. As already observed in  \cite{Donagi:2010pd,Marsano:2011nn,Kerstan:2012cy}, this condition allows us to understand  the M5-analogue of a D3-instanton in which charged chiral zero modes are induced only by 7-brane worldvolume fluxes. In this case, the instanton generates terms which include operators charged under $U(1)$ gauge symmetries that are massive by means of a flux induced St\"uckelberg mechanism. We will come back to this effect in section \ref{sec:T3=0}.
  
 However, we will see that in fact  (\ref{T3BI}) contains considerably more information than just its cohomological counterpart (\ref{cohoBI}), in particular in relation with possible hidden massive $U(1)$s. This will clarify the connection with the weakly coupled IIB picture. Roughly, we can think of $T_3$  as a smeared M2-brane instanton wrapping
 a 3-chain $\Gamma\subset D$ that ends on the 2-cycle  $C_G-C$, where $C_G\subset D$ is a curve which is Poincar\'e dual to $G_4|_D$. This allows us to link the present setting to the discussion of M2-brane instantons of the companion paper \cite{Martucci:2015dxa}, in which it is shown how the M2-instanton generated couplings depend on some refined information on the 3-chain $\Gamma$, and not just its homological characterisation. Hence, we expect something similar to happen for the $T_3$-flux on the M5-instanton. We will come back to this idea and explore it  in great detail starting from section \ref{sec_pertvsnonpert}.  Here we first focus on a crucial difference between the (localised) M2-instantons  and the  (delocalised) $T_3$-flux. Namely, M2-brane instantons on Calabi-Yau 4-folds always break supersymmetry and can only generate four-dimensional D-terms \cite{Martucci:2015dxa} while, as discussed in the following subsection, the $T_3$-flux can in principle preserve supersymmetry so that the fluxed M5-instanton can generate F-terms in the four-dimensional effective theory.

A final couple of remarks regarding  (\ref{T3BI}). First, recall that the $G_4$-flux is not always integrally quantised because of  the bulk quantisation condition $G_4-\frac12\lambda\in H^4(\hat Y,\mathbb{Z})$, which contains the potentially half-integer shift  $\frac12 \lambda=\frac14 p_1(\hat Y)\equiv - \frac12 c_2({\hat Y})$  \cite{Witten:1996md}. On the other hand, consistency with (\ref{T3BI}) and (\ref{cohoBI}) requires such a possible half-integer shift to cancel once $G_4$ is restricted to the divisor $D$ \cite{Witten:1996hc,Witten:1999vg}. Let us explicitly check this, assuming  that the effective divisor $D$ is smooth, as it is \emph{generically} the case in our setting once $\Y$ is resolved into $\hat Y$ (see, however, the discussion in section \ref{sec:gaugenh}).  The  sequence of complex bundles
\be
0\ \rightarrow\  N_D\ \rightarrow\  T_{\hat Y}|_D\  \rightarrow\  T_D\ \rightarrow\  0
\ee
implies the adjunction formula for Chern classes, $c(\hat Y)|_D=c(D) c(N_D)$. Hence, by using $c_1(\hat Y)=0$, one obtains  $c_1(N_D)=-c_1(D)$ and 
$c_2(\hat Y)|_D=c_2(D)-c_1(D)^2$, so that  we can rewrite $\lambda|_D$ just in terms of Chern classes on $D$,
\be
\lambda|_D=c_1(D)^2-c_2(D).
\ee
On the other hand, as in \cite{Collinucci:2010gz}, one can prove that  $c_1(D)^2-c_2(D)$  is an {\em even} integer cohomology class, so that $\frac12\lambda|_D\in H^4(D,\mathbb{Z})$.

Second, we have not explicitly written a possible anomalous torsional contributions to (\ref{T3BI}) \cite{Witten:1999vg,Diaconescu:2003bm,Belov:2006jd},  implicitly reabsorbing it in a redefinition of $G_4|_D$. If non-vanishing, such torsional contribution should be cancelled by appropriate M2-brane insertions and in the following we assume that such cancellation is  possible  just by means of fibral curves.

\subsection{Supersymmetry}
\label{sec:susy}

We are primarily interested in supersymmetric M5-brane instantons. The M-theory background has the structure described in \cite{Becker:1996gj}, which reduces in the F-theory limit to a type IIB background as in \cite{GKP}. To the best of our knowledge, the M5-brane supersymmetry conditions on such backgrounds have never been systematically derived so far. However, as in \cite{Bianchi:2011qh,Bianchi:2012kt,Martucci:2014ema}, one can use the IIB  perspective, in which such instantons are given by Euclidean D3-branes that wrap four-cycles $D_{\rm b}$ in the base $\B$ of the elliptically fibered 4-fold $\Y$ and possibly support a non-trivial self-dual world-volume flux $\calf=\frac{1}{2\pi}F_{E}-\iota^* B_2$. Supersymmetry then requires that $D_{\rm b} \subset \B$ is holomorphically embedded and that $\calf$ is self-dual, $*\calf=\calf$, i.e.\ $\calf$ is purely $(1,1)$ and primitive. The corresponding M5-brane wraps  the vertical divisor $D = \pi^{-1} D_{\rm b}$ inside $\Y$,  as already assumed above. Furthermore, the  D3-brane flux $\calf$ naturally uplifts to a $(2,1)$ and primitive $T_3$ flux on the M5-brane \cite{Bianchi:2012kt}, which is indeed compatible with the imaginary-self-duality condition (\ref{T3ISD}). We will assume these as the supersymmetry conditions to be imposed on $T_3$, i.e.
\begin{subequations}\label{T3susy}
\begin{align}
T_3&\equiv T_{2,1}, \label{T3susy1}\\
 J_D&\wedge T_3=0 \label{T3susy2},
\end{align}
\end{subequations}
where $J_D\equiv J|_D$.
Notice that the supersymmetry condition (\ref{T3susy1})  appears to be inconsistent for a real and properly quantised $T_3$ satisfying the Bianchi identity (\ref{T3BI}).  This problem actually shows up, as is well-known, already at the level of the imaginary-self-duality condition (\ref{T3ISD}). One way  to properly address this issue is to treat $T_3$ as an unconstrained quantised real field and impose the conditions (\ref{T3susy}) and (\ref{T3ISD}) at the partition function level as in \cite{Witten:1996hc}, see for instance \cite{Witten:2007ct} for a brief non-technical introduction to this approach and references therein for more details. 

Let us first assume that the elliptic fibration $Y$ can be crepantly resolved into a smooth Calabi-Yau four-fold $\hat \Y$. Correspondingly, the vertical divisor $D$ wrapped by the M5-brane is also generically resolved into a smooth K\"ahler elliptic fibration  -- see section, however, \ref{sec:gaugenh} for non-generic situations  -- which we will denote by the same symbol  $D$. Under such resolution the fibral curves $C_A$ entering the combination $C=\sum_A n^A C_A$ in (\ref{T3BI}) acquire a finite volume 
\be\label{posvol}
{\rm vol}_2(C_A)=\int_{C_A} J>0.
\ee
Consider now an unconstrained real $T_3$ satisfying the Bianchi identity (\ref{T3BI}). This can always be solved by Hodge-decomposing  $T_3$ as 
\be\label{T3dec}
T_3=T^{\rm harm}_3+\d\alpha_2+\d^\dagger\gamma_4\,.
\ee
Notice that $\gamma_4$ is defined up to a co-closed contribution. This freedom can be partially fixed by imposing the condition $\d\gamma_4=0$. Hence $(\ref{T3BI})$ reduces to $\Delta\gamma_4=G_4|_{D}-\delta_4(C)$. Once the cohomological condition (\ref{cohoBI}) is satisfied, this equation has the solution
\be\label{gamma4dec}
\gamma_4=\gamma^{\rm harm}_4+\calg[G_4|_D - \delta_{4} (C)],
\ee
where $\calg$ is the Green's operator on $ D$. One can regard (\ref{T3dec}) together with (\ref{gamma4dec}) as a parametrisation of the off-shell unconstrained $T_3$ satisfying (\ref{T3BI}).  In particular, 
\be\label{zeroT3}
T_3^{(0)}=\d^\dagger\calg[G_4|_D - \delta_{4} ({\cal C})]
\ee
 represents the `background' world-volume flux produced by the $G_4$-flux and the incoming M2-brane insertions, $T^{\rm harm}_3$ distinguishes different topological sectors, while $\alpha_2$ represents the dynamical degrees of freedom.  

It is now easy to see that the supersymmetry condition (\ref{T3susy2}), which is perfectly compatible with a real and quantised $T_3$, cannot be satisfied for a resolved ambient space $\hat Y$. Indeed, by taking into account the bulk supersymmetry condition $J\wedge G_4=0$, plugging (\ref{T3dec}) and (\ref{gamma4dec}) into (\ref{T3susy2}) one gets
\be\label{primdec}
J_D\wedge T^{\rm harm}_3+\d(J_D\wedge \alpha_2)-\d^\dagger\calg[J_D\wedge\delta_{4} (C)]=0,
\ee
where we have also used the fact that $J_D$ commutes with the operators $\d$, $\d^\dagger$ and $\calg$. Since $J_D\wedge T^{\rm harm}_3$ is harmonic, by the uniqueness of the Hodge decomposition (\ref{primdec}) can be satisfied only if each of its contributions separately vanishes. In particular, the vanishing of the third contribution is possible only if \footnote{Notice that $J_D\wedge\delta_{4} (C)$ is exact by (\ref{cohoBI}) and $J\wedge G_4=0$. Then,  by applying $\d$ to  $\d^\dagger\calg[J_D\wedge\delta_{4} (C)]=0$ we get $0=\d\d^\dagger\calg[J_D\wedge\delta_{4} (C)]=\Delta\calg[J_D\wedge\delta_{4} (C)]=J_D\wedge\delta_{4} (C)$. By writing $\delta_{4} (C)=\sum_an^a\delta_{4} (C_a)$ one obtains (\ref{vanishingvolcond}).}
\be\label{vanishingvolcond}
J_D\wedge\delta_{4} (C_A)=0
\ee
for any fibral curve $C_A$ appearing in $C=\sum_A n^A C_A$. On a resolved $\hat Y$ (\ref{vanishingvolcond}) is not satisfied because of (\ref{posvol}) and so the supersymmetry condition (\ref{T3susy2}) can be satisfied only in the F-theory limit. In fact, the appearance of such obstruction in the resolved phase is expected. Indeed, in this case the M2-branes attached to the M5-brane are massive and the M5 is expected to bend under their tension, as for instance in the local solution described in \cite{Howe:1997ue}. Hence in this regime our supersymmetry conditions, which assume the Euclidean M5-brane  to be localised at a certain external space-time point, cannot be accurate. On the other hand,  in the F-theory limit ${\rm vol}_2(C_A)\rightarrow 0$ and such obstruction disappears. Hence, even though we have initially assumed the crepant resolvability of the ambient space $Y$, supersymmetric M5-brane instantons with M2-brane insertions of the form considered here  are really consistent only in the limit of vanishing volume for the fibral curves $C_A$. This is just sufficient for our purposes as we are interested in taking the F-theory limit after all. It is then natural to extend our conclusions to the case in which $Y$ does not admit a crepant resolution.  
 
The above discussion implies that the primitivity condition reduces to requiring that $J_D\wedge \alpha_2$ is closed and that $J_D\wedge T^{\rm harm}_3=0$. The first condition can be always satisfied by appropriately choosing $\alpha_2$. Hence, the primitivity condition is equivalent to the cohomological condition
\be\label{cohoprim}
[J_D\wedge T_3]=0\quad \text{in $H^5(D)$}.
\ee
 
As already mentioned, while the primitivity condition (\ref{T3susy2}) can be studied at the classical level on an unconstrained real $T_3$, the other supersymmetry condition (\ref{T3susy1}) cannot. To proceed as in \cite{Witten:1996hc}, one should decompose $T_3=T_3^++T^-_3$, with $*T_3^\pm=\pm\ii T^+_3$, and isolate the contribution of $T^+_3$ to the partition function. This important step is beyond the scope of the present paper and we will not try to attack it here. However notice that, once the primitivity condition (\ref{T3susy2}) is imposed on the unconstrained $T_3$, the remaining condition (\ref{T3susy1}) can be satisfied by its self-dual component $T_3^+$ if and only if $T_{0,3}=0$ (and then $T_{3,0}=0$ too) for the unconstrained $T_3$. In particular the contribution (\ref{zeroT3})  has vanishing $(0,3)$ component by construction  so that it is automatically compatible with the condition $T_{0,3}=0$. In other words, (\ref{T3BI}) implies that $\del T_{0,3}=0$, so that $T_{0,3}$ defines a class $[T_{0,3}]\in H^{0,3}(D)$ and the supersymmetry condition $T_{0,3}$  can be satisfied if and only if the cohomological condition 
\be\label{cohohol}
[T_{0,3}]=0\quad \text{in $H^{0,3}(D)$}
\ee
is satisfied.

To summarise, we see that in the F-theory limit the supersymmetry conditions (\ref{T3susy}) are equivalent, at the level of the unconstrained flux $T_3$, to the cohomological conditions (\ref{cohoprim}) and (\ref{cohohol}). Once such conditions are satisfied, there is no obstruction for the M5-brane instanton to be supersymmetric. Interpreting $T_3$ as a Euclidean M2-brane dissolved into the M5-brane, we see how such dissolution allows the M2-M5-bound state to be supersymmetric, while an M2-brane instanton alone could never be supersymmetric in F-theory compactifications to four dimensions \cite{Martucci:2015dxa}.

An interesting observation is that the condition $T_{0,3}=0$ for the unconstrained $T_3$  and the condition on the holomorphy of the vertical 6-cycle $D$, which may be regarded as  `holomorphy' conditions for the supersymmetric M5-brane, can be both derived from a `superpotential'
\be\label{T3funct}
W[D,T_3]=\int_\Theta\Omega_\Y\wedge T_3.
\ee  
Here $\Omega_\Y$ is the holomorphic 4-form on $\Y$ and  $\Theta$ is a 7-chain interpolating between a general off-shell 6-cycle $D$ and  a reference 6-cycle $D_0$, $\del\Theta=D-D_0$. Since $D$ and $D_0$ are vertical 6-cycles, the 7-chain $\Theta$ can be regarded as an elliptic fibration over a 5-chain in the base. Furthermore, we are assuming that $T_3$ can be extended onto the 7-chain $\Theta$ and that the fibral curves $C_a$ wrapped by the M2-branes insertions follow the interpolation between $D_0$ into $D$ by spanning  a 3-chain $\Gamma\subset \Theta$ with $\del\Gamma\subset\del\Theta$. The functional (\ref{T3funct}) was partially anticipated in \cite{Kerstan:2012cy} and is completely analogous to the superpotentials for D-branes derived in \cite{luca2}, which generalise the superpotential for D5-branes on 2-cycles of \cite{Witten:1997ep}.\footnote{Notice that (\ref{T3funct}) must be well-defined, in the sense that it must not depend on the specific choice of the 7-chain $\Theta$. This can indeed be checked by appropriately extending the  Bianchi identity (\ref{T3BI}) and the 3-chain $\Gamma$ away from a given 7-chain $\Theta$.}   

It is  easy to see how (\ref{T3funct}) produces the holomorphy conditions. The extremisation with respect to the embedding moduli gives $(\iota_\zeta\Omega_Y)|_D\wedge T_3=0$, for any vector $\zeta\in TY|_D$, which is equivalent to the condition $T_{0,3}=0$. On the other hand, by extremising  with respect to $\delta T_3=\d\delta\alpha_2$ one gets   $(\Omega_Y\wedge \delta\alpha_2)|_D=0$,  which is equivalent to requiring that $D$ is holomorphic.


\section{Perturbative vs non-perturbative homology on the M5 and $G_4$-flux} \label{sec_pertvsnonpert}

Since our main strategy is to deduce instanton generated couplings based on a careful analysis of the Bianchi identity (\ref{T3BI}) we need to discuss in more detail the structure of the vertical divisor $D$ wrapped by the M5-brane.
The world-volume $D = \pi^{-1} D_{\rm b}$ is a (non-Calabi-Yau) elliptically fibered 3-fold itself.  
The elliptic fibre of $D$ degenerates over the loci on the base where $D_{\rm b}$ intersects the discriminant locus of $\Y$. 
Note that $D_{\rm b}$ generically intersects the locus of the 7-branes on $B$ in a complex curve and the matter curves in points.\footnote{If, by contrast, the divisor $D_{\rm b}$ is contained in the discriminant, the M5-instanton corresponds to a gauge instanton or even to the F-theory analogue of the more exotic type of configurations studied in \cite{Petersson:2007sc}. We will not discuss these special cases in this paper.}
In particular, by going to a resolved ambient space $\hat\Y$, we can unambiguously
identify those 2-cycles which are generated by  fibral curves over such codimension-one or -two loci on $D_{\rm b}$. 
For these fibral curves on $D$ we can distinguish between {\em perturbative} and {\em non-perturbative} homological relations in the sense introduced in \cite{Martucci:2015dxa}.

More precisely, consider a linear combination $C=\sum_A n^A C_A$ of fibral curves $C_A$ in $D$, sitting at (possibly different) base points in $D_{\rm b}$, and suppose that such combination is trivial in $H_2(D,\mathbb{Z})$,
\be\label{Dhom}
[C]_D=0.
\ee
If we can continuously move the $C_A$ along $D$ to the same base point $p$ while preserving holomorphicity and
\be \label{homrelba}
\sum_A n^A \,[C_A] = 0 \in H_2(\mathfrak f_p, {\mathbb Z})
\ee
in the elliptic fibre $\mathfrak f_p$ over $p$,  we call the homological relation (\ref{homrelba}) {\em perturbative}. In other terms, there exists a 3-chain  $\Gamma$ trivialising $C$ in $D$, $\del\Gamma=-C$, with vanishing volume in the F-theory limit. For this reason, we will refer to it as a {\em perturbative} 3-chain and usually denote it by $\Gamma_{\rm p}$.  
 In this case, by adapting the notation introduced in \cite{Martucci:2015dxa},  we write
\be
[C]_{D,{\rm p}}=0\, .
\ee
Furthermore, we say that $C$ is perturbatively equivalent to $C'$ if 
\be
[C-C']_{D,{\rm p}}=0\,.
\ee
Now, one could have $[C]_D=0$ but $[C]_{D,{\rm p}}\neq 0$. Then we say that $C$  is {\em non-perturbatively} trivial and we write
\be
[C]_{D,{\rm np}}=0\,.
\ee 
In this case any 3-chain  trivialising $C$ in $D$ has non-vanishing volume in the F-theory limit; it will be referred to as {\em non-perturbative} and usually denoted by $\Gamma_{\rm np}$.

We also recall that we use the notation $[C]_{{\rm p}}=0$  (or $[C]_{{\rm np}}=0$) for a fibral curve $C\subset \hat\Y$ that is perturbatively (or non-perturbatively) trivial  in the bulk $\hat\Y$. 
Notice that $[C]_{D,{\rm p}}=0$ on the M5-brane implies that $C$ is perturbatively trivial in the bulk of $\hat Y$, $[C]_{{\rm p}}=0$. The reversed direction is more involved:
\begin{enumerate}
\item
The non-perturbative homological relation $[C]_{{\rm np}}=0$ on $\hat \Y$ implies either $[C]_{D,{\rm np}}=0$ or $[C]_{D}\neq 0$ on the M5-brane.

To see how the latter possibility can arise recall that the vertical divisor $D = \pi^{-1}D_{\rm b}$ is by itself a non-trivially elliptically fibered 3-fold (albeit not a Calabi-Yau). By the Shioda-Tate-Wazir theorem, the group of divisors on $D$ is generated by the pullback of divisors from the base $D_{\rm b}$, the resolution divisors of non-abelian singularities in codimension-one on $D$ as well as the group of sections. 
A  fibral curve is homologically trivial, up to torsion,  if and only if it has zero intersection number with each resolution divisor or section.
These first include the pullback of the resolutions divisors\footnote{If $D_{\rm b}$ misses part of the discriminant locus, some of the resolution divisors pullback trivially to $D$.} and sections of the ambient 4-fold $\hat Y$. 
Now, a fibral curve with $[C]_{{\rm np}}=0$ has vanishing intersection with the resolution divisors and the sections on $\hat Y$, and this remains true for the inherited intersection product with pullback resolution divisors and sections on $D$.
However, $D$ might exhibit extra such divisors which do not arise by pullback from $\hat Y$. In this case, $[C]_{{\rm np}}=0$ need not imply $[C]_{D} = 0$.
This necessitates that $D$ is singular as an elliptic fibration even though this singularity is not a singularity of the 4-fold $\hat Y$. Since the singularity sits in the fibre, the base $D_{\rm b}$ can be perfectly smooth, and therefore the contribution of such instantons  to the partition function can be treated in the usual way, up to a suitable resolution of the fibre singularity.

On the other hand, from this discussion it should be clear  that, {\em generically}, $[C]_{{\rm np}}=0$ on $\hat \Y$ implies $[C]_{D,{\rm np}}=0$ on $D$. Concrete examples in which this generic expectation is violated are discussed in section \ref{sec:gaugenh} and are related to a `$U(1)$ enhancement' along the M5-brane instanton.

\item
 $[C]_{{\rm p}}=0$ in the bulk can correspond to $[C]_{D,{\rm p}}=0$ or to  $[C]_{D,{\rm np}}=0$ or even to $[C]_{D}\neq 0$ on the M5-brane. 
 
The first possibility occurs for instance if the perturbative equivalence $[C]_{{\rm p}}=0$ in the bulk can be satisfied fibrewise so that the restriction to $D$ does not affect this property. 
On the other hand, the second option is generically expected to occur when the perturbative equivalence $[C]_{{\rm p}}=0$ in the bulk involves non-trivial paths (e.g.\ monodromies\footnote{ An example of such a monodromy, as explained in section 5 of \cite{Martucci:2015dxa}, is the monodromy for the $\mathbb Z_2$ charged matter states in the bisection fibration discussed in \cite{Braun:2014oya,Morrison:2014era,Anderson:2014yva,Klevers:2014bqa,Garcia-Etxebarria:2014qua,Mayrhofer:2014haa,Mayrhofer:2014laa}. See \cite{Cvetic:2015moa} for the discussion of a trisection model.}.)   of the fibral curves $C_A$ appearing in $C=\sum_A n^A C_A$ along matter curves in the base -- see \cite{Martucci:2015dxa} for illustrative examples. Indeed, such paths are generically obstructed by the restriction to $D$ and then,  in the generic case in which $D$ does not exhibit extra  divisors, one has  $[C]_{D,{\rm np}}=0$ by the same arguments as in point 1.   In section \ref{sec:G4trivial}, starting from this observation we will discuss a mechanism for  generating non-perturbatively supressed corrections to the rank-one perturbative Yukawa couplings.
 The last possibility, $[C]_{{\rm p}}=0$ but $[C]_{D}\neq 0$, albeit non-generic, can still occur  similarly  to what has been discussed under point 1. 
\end{enumerate}

We would like to reinterpret the Bianchi identity (\ref{T3BI}) in Poincar\'e dual terms of this refined version of homology. To this aim we need to use a description of the $G_4$ flux which involves holomorphic cycles. This is naturally achieved in the description of $G_4$ flux by algebraic cycles \cite{Braun:2011zm}. According to the prescription of \cite{Bies:2014sra}, a $G_4$ flux can be specified by an algebraic `transversal' 4-cycle $\cald_G\in \hat \Y$, up to rational equivalence. 
 Recall that two algebraic cycles $\gamma_1$ and $\gamma_2$ are rationally equivalent if there is a family of algebraic cycles $\Gamma_t\subset \hat Y$ that is rationally parametrised by $t \in \mathbb P^1$ such that $\Gamma_{t_1}=\gamma_1 $ and  $ \Gamma_{t_2}=\gamma_2 $ -- see for instance \cite{EisenbudHarris} for a more complete definition. This is a generalisation of the notion of linear equivalence for divisors. The importance of rational equivalence comes from the fact that 
the fluxes associated with two rationally equivalent 4-cycles describe the same background configuration of the M-theory 3-form $C_3$ and its field strength, up to gauge equivalence. The rational equivalence class of such 4-cycles is denoted by ${\rm CH}^2(\hat \Y)$, the second Chow group on $\hat \Y$.\footnote{More precisely, it is the D\'eligne cohomology group $H^4_{\cal D}(\hat Y, \mathbb Z(2))$ \cite{EsnaultDeligne} which counts the 3-form background up to gauge equivalence \cite{Curio:1998bva} (see also \cite{Clingher:2012rg,Anderson:2013rka}). A refined cycle map from ${\rm CH}^2(\hat \Y)$ to $H^4_{\cal D}(\hat Y, \mathbb Z(2))$ allows us then to parametrise elements of $H^4_{\cal D}(\hat Y, \mathbb Z(2))$ by algebraic cycles up to rational equivalence \cite{Bies:2014sra}. The superscript in ${\rm CH}^2(\hat Y_4)$ denotes the complex codimension of the algebraic cycles in question.}

The  restriction of $G_4$ to the M5-brane world-volume can be identified with a `transversal' curve $C_G$ on $D$,
again defined up to rational equivalence inside $D$. These in turn are elements of ${\rm CH}^{2}(D)$.\footnote{The important point is that the inclusion $\iota: D \rightarrow \hat Y$ induces a well-defined pullback $\iota^*: {\rm CH}^2(\hat Y) \rightarrow {\rm CH}^2(D)$. Via this map rationally equivalent 4-cycles on $\hat Y$ define rationally equivalent 2-cycles on $D$. \label{footnoteChow1}} Furthermore, the transversality condition implies that $C_G$ must be a linear combination of fibral curves, $C_G=\sum_A n_G^A \,  C_A$. We can then classify $C_G$ according to our refined definition of perturbative homology.  Indeed rational equivalence implies perturbative homological equivalence.
Thus the perturbative homology class of a certain fibral curve is invariant under deformations preserving the rational equivalence class. It makes therefore sense to speak of perturbative or non-perturbative homological relations for rational equivalence classes of curves.
This is important because the $G_4$-flux is defined only up to rational equivalence, and perturbative or non-perturbative homological relations are well-defined concepts for $G_4$.

 Suppose first that $[C_G]_D=0$ in ordinary homology. In this case we could  have  either $[C_G]_{D,{\rm p}}=0$ or  $[C_G]_{D,{\rm p}}\neq 0$ (and then just $[C_G]_{D,{\rm np}}=0$). 
On the other hand, if $[C_G]_D\neq 0$ in ordinary homology, then necessarily $[C_G]_{D,{\rm np}}\neq 0$ and thus also  $[C_G]_{D,{\rm p}}\neq 0$. As a result, perturbative homology still allows us to distinguish between, say,  $C_G$ and $C'_G$ which are homologous. Indeed, even if $[C_G-C'_G]_{D}= 0$, we could have $[C_G-C'_G]_{D,{\rm p}}\neq 0$ and  $[C_G-C'_G]_{D,{\rm np}}= 0$, so that $C_G$ and $C'_G$ can be identified only at the non-perturbative level.

We are now in the position to discuss the implications of this refined homological description of fibral curves and $G_4$ fluxes for the structure of F-terms produced by an M5-brane instanton.

\section{M5 selection rules: first case ($G_4|_D= 0$)}
\label{sec:G4trivial}

In this section we consider M5-instantons such that $G_4|_D  \equiv 0$ as is for instance the case  if  $G_4\equiv 0$ in the bulk. 
Actually we assume that $C_G= 0$ modulo rational equivalence, which means that we are actually setting to zero the restriction to $D$ of possible background Wilson lines. 
The Bianchi identity (\ref{T3BI}) becomes 
\be\label{BInoflux}
\d T_3 = -\delta_{4}(C)\,.
\ee
In this case, $C$ must be  homologically trivial both in $D$ and in the bulk and   matter  states associated with  M2-branes on $C$ 
cannot carry any net charge under a massless or discrete gauge symmetry. 

More precisely, in order to satisfy (\ref{BInoflux}) we can either have $[C]_{D,{\rm p}}=0$ or $[C]_{D,{\rm np}}=0$. In the first case, $[C]_{D,{\rm p}}=0$, we can in fact set $\delta_{4}(C)=0$ in (\ref{BInoflux}), which then reduces to $\d T_3=0$. In this case the M5-instanton and the M2-branes decouple: the M2-brane can generate an independent perturbative term in the effective action while the M5-brane instanton can generate a term which does not contain charged matter, as in the smooth case originally considered in \cite{Witten:1996bn}.

 The case $[C]_{D,{\rm np}}=0$ is conceptually more interesting. In this case there is a non-perturbative 3-chain $\Gamma_{\rm np}\subset D$ with 
\be
\partial \Gamma_{\rm np} =- C.
\ee
 This implies that, in order to solve (\ref{BInoflux}),  a suitable world-volume flux $T_3$ must be turned on. 
Notice that this information is lost in the purely cohomological   counterpart of (\ref{BInoflux}), which  is just given by $[C]_D=0$.

Such a fluxed M5-brane instanton admits an interpretation as a bound state of an M5-instanton with an M2-instanton wrapping the 3-chain $\Gamma_{\rm np}$ as considered in \cite{Martucci:2015dxa}.\footnote{Clearly, in such correspondence, changing  $\Gamma_{\rm np}$ by a closed  3-cycle corresponds to changing the harmonic component of  $T_3$.} The formation of a bound state out of the M5- and the M2-instanton requires the M2-charge to spread out into a delocalised $T_3$.   The key point is that, while  an M2-brane instanton wrapping $\Gamma_{\rm np}$ would necessarily break supersymmetry, as explained in section \ref{sec:susy} this is not anymore true  for its delocalised counterpart given by $T_3$.   Hence, such fluxed M5-branes can generate F-terms. The more specific kind of  produced F-term, and in particular whether or not a superpotential is generated, depends on the structure of neutral fermionic zero modes  \cite{Witten:1996bn}, which is generically affected by the world-volume flux itself \cite{Bianchi:2011qh,Bianchi:2012kt}.  We will not consider such details in the sequel, assuming for simplicity  that a superpotential is generated  and focusing  on its dependence on charged matter. 

If $C=\sum_A n^A C_A$ as above and we denote the state associated with an M2-brane on $C_A$ by $\Phi_A$, then the fluxed M5-instanton generates a superpotential of the schematic form
\be\label{chargeW}
W\simeq \prod_A (\Phi_A)^{n^A} \, e^{ - S_{\rm M5}}
\ee
in the effective action. The prefactor $\prod_A (\Phi_A)^{n^A}$ corresponds to the insertion of the time-like M2-branes ending on the M5-brane instanton. Notice that such prefactor is actually not unique, but is defined  up to  insertions of possible perturbatively allowed operators, which correspond
to changing $C$ by 2-cycles that are perturbatively trivial. 

Under a gauge transformation $C_3 \rightarrow C_3 + \d \Lambda_2$, from (\ref{M2action}) it is clear that the M2-brane insertion in the path-integral transforms as 
\be\label{expM2}
e^{-S_{\rm M2}}\quad\rightarrow\quad e^{2\pi\ii \int_C \Lambda_2}\,e^{-S_{\rm M2}}
\ee
and then, correspondingly, we can conclude that
\be
\prod_A (\Phi_A)^{n^A}\quad\rightarrow\quad e^{2\pi\ii \int_\calc \Lambda_2}\,\prod_A (\Phi_A)^{n^A}\,.
\ee
This shows that the operator $\prod_A (\Phi_A)^{n^A}$ is charged under a gauge transformation with parameter  $\int_C \Lambda_2$. 
 $C$ being trivial in homology,  such gauge transformation can be attributed to a massive four-dimensional $U(1)$ gauge symmetry in the sense of \cite{Grimm:2011tb,Braun:2014nva,Martucci:2015dxa}.  On the other hand, by using (\ref{BInoflux}), under such gauge transformations we also have
\be\label{expM5}
e^{-S_{\rm M5}}\quad\rightarrow\quad e^{-2\pi\ii \int_C \Lambda_2}\,e^{-S_{\rm M5}}\,,
\ee
see  (\ref{completeM5}) and  (\ref{M5action}). Hence the superpotential (\ref{chargeW}) is  gauge invariant, while (\ref{expM2}) and (\ref{expM5}) are not separately consistent. 

We then conclude that a superpotential of the form (\ref{chargeW}), and more generically an F-term with a similar structure, can be generated, which is exponentially suppressed by $e^{-2\pi{\rm vol}(D)}$. This has to be contrasted with the fact that a perturbative (unsuppressed)  superpotential term of the form $\prod_A (\Phi_A)^{n^A}$  is not allowed as a result of a massive $U(1)$ gauge symmetry \cite{Martucci:2015dxa}.

We would now like to connect this purely M/F-theoretical discussion to its weakly-coupled  IIB counterpart. 
Consider a D3-instanton supporting a   flux $F_E \in H^{2}_-(E)$.
 If $q_a=\frac1{2\pi}\int_{\gamma_a} F_E=0$ for any intersection curve  $\gamma_a=E\cap D_a$ between the D3-brane and the D7-brane, there are no net chiral fermionic zero modes on $\gamma_a$ -- see equation  (\ref{chiralindexa}) -- and then 
there is no topological obstruction to the contribution of such D3-instanton to the partition function. In this case $F_E$ uplifts to a closed 3-form $T_3$ on the M5-worldvolume $D$. 

By contrast, if $\int_{\gamma_a} F_E\neq 0$ for some $\gamma_a$, then $\gamma_a$ supports a net number of chiral charged instanton zero modes via (\ref{chiralindexa}). This prevents  the D3-instanton 
from contributing to the path-integral without further insertions. Correspondingly, there is an obstruction to uplifting  $F_E$ to a well-defined $T_3$ on the corresponding M5-brane.  
The rationale behind this obstruction is Poincar\'e dual to what  we have discussed for D1-instantons in \cite{Martucci:2015dxa}. To see this
we consider the orientifold-odd 2-cycle $\Sigma_E$ dual to $F_E$ on $E$ and notice that $\int_{\gamma_a} F_E$ counts the net number $\Sigma_E\cdot \gamma_a $ of intersection points between $\Sigma_E$ and $\gamma_a$. Since a D7-brane on $D_a$ induces a $T$-monodromy in the fibre around $\gamma_a$, this monodromy obstructs the F-theory uplift of $\Sigma_E$ to a well defined 3-cycle in $D$, which should represent the Poincar\'e dual of a closed $T_3$. A consistent configuration then requires that this monodromy must be cancelled by an opposite intersection with another 7-brane $D_b$, such that the curve $\gamma_b=E\cap D_b$ intersects $\gamma_b$ at the set of points $p_{ab}$. 
The analogy with the discussion in \cite{Martucci:2015dxa}  suggests  identifying the non-holomorphic odd 2-cycle $\Sigma_E$ with the curve wrapped by a D1-instanton. However, in that case $\Sigma_E$ 
would have to intersect $\gamma_a$ and $\gamma_b$ with opposite intersection number by passing through one of the  points $p_{ab}$. 
For the D3-instanton flux $F_E$ dual to $\Sigma_E$, this localisation requirement  can be relaxed due to the dilution of the flux and the only condition is that  $q_a=-q_b=n$, with  $q_b=\int_{\gamma_b} F_E$.
In this way the net number of $\eta_a$ and $\tilde\eta_b$ zero modes can be the same and the D3-brane partition function can generate a non-trivial term of the kind (\ref{Phi-coup1a}) in the effective action.
Correspondingly, the D3-instanton flux $F_E$ with such intersection numbers uplifts to a non-closed 3-form $T_3$ satisfying the Bianchi identity (\ref{BInoflux}) which is the dual of a non-perturbative 3-chain $\Gamma_{\rm np}$ ending on the fibral curve $C$. In this way, as we discussed above, one can insert time-like M2-branes wrapping $C$, getting a consistent configuration. This  corresponds to the absorption of the D3 zero-modes  by interaction terms  (\ref{intterms}) and produces the prefactor $(\Phi_{ab})^n$ in (\ref{Phi-coup1a}).

\subsection{M5/D3 instantons in the massive $U(1)$ model}

As an example consider the massive $U(1)$ model introduced in \cite{Braun:2014nva}.
We have discussed this model from the perspective of D1/M2-instanton generated operators in section 6 of our companion paper \cite{Martucci:2015dxa}, to which we refer for further details. In the weakly coupled IIB description, there are two D7-branes along divisors $D_1$ and $D_2$ on a Calabi-Yau 3-fold $X$. They intersect along a curve $\calc_{12} = D_1 \cap D_2$  away from the O7-plane where a chiral multiplet $\Phi$ charged only under a massive $U(1)$ group is localised. The curve $\calc_{12}\subset \X$  and its orientifold image $\calc'_{12}$ are the uplift of a curve in the base $\B$, denoted as $\calc_{II}$ in \cite{Martucci:2015dxa}, over which the elliptic fibre  degenerates to an $I_2$-fibre. Once resolved this fibre contains a resolution $\mathbb P^1_E$ which is homologically trivial \cite{Braun:2014nva}, but only in a non-perturbative sense, that is,   $[\mathbb P^1_E]_{\rm np}=0$ but  $[\mathbb P^1_E]_{\rm p}\neq 0$   \cite{Martucci:2015dxa}. Hence it is capped by a non-perturbative 3-chain $\Gamma_{\rm np}$ whose volume does not vanish  in the F-theory limit. This 3-chain uplifts an orientifold-odd curve $\Sigma \in H_{2}^-(X)$ intersecting $D_1$ and $D_2$ with intersection numbers $+1$ and $-1$, respectively \cite{Martucci:2015dxa}. A D1-instanton on $\Sigma$ uplifts to an M2-instanton on $\Gamma_{\rm np}$. Both can generate a D-term of the schematic form $\Phi \, e^{-S}$.

Consider now an M5-instanton along a divisor $D = \pi^{-1} D_{\rm b}$ and suppose that the base divisor $D_{\rm b}$ intersects the matter curve $C_{II}$ in a number of isolated points $p_k$. Over each point $p_k$ sits a copy  $\mathbb P^1_{E,k}$ of the fibral resolution curve. 
In the language of section \ref{sec_pertvsnonpert}, for two different points $p_k \neq p_l$, even though  $\mathbb P^1_{E,k}$ and  $\mathbb P^1_{E,l}$ are perturbatively homologous in the bulk (that is, $[\mathbb P^1_{E,k} - \mathbb P^1_{E,l}]_{{\rm p}} = 0$), they are not on the M5-brane,
\be \label{P1kP1l}
[\mathbb P^1_{E,k} - \mathbb P^1_{E,l}]_{D, {\rm p}} \neq 0 .
\ee
On the other hand, as discussed in section \ref{sec_pertvsnonpert},  for a generic vertical divisor $D$, if a fibral curve on $D$ is non-perturbatively trivial in the bulk then it is so 
also on $D$. Hence we generically have  $[\mathbb P^1_{E,k}]_{D, {\rm np}} = 0$ and then $D$ contains a 3-chain $\Gamma_{\rm np}$ bounded by, say $\mathbb P^1_{E,1}$, with non-vanishing volume in the F-theory limit. Hence, by inserting a time-like M2-brane wrapping $\mathbb P^1_{E,1}$ and turning on a non-closed $T_3$ flux Poincar\'e dual to $\Gamma_{\rm np}$ one obtains a consistent  M5-brane instanton, which  can  then generate an F-term  coupling $\Phi \, e^{- S_{M5}}$.

 At weak coupling, such contribution corresponds to a D3-instanton wrapping a divisor $E\subset \X$ and supporting a flux $F_E$ with $\int_{\gamma_1} F_E=-\int_{\gamma_2} F_E=1$, where $\gamma_1=D_1\cap E$  and $\gamma_2=D_2\cap E$.  
 An example of such flux configuration can be one for which the individual numbers (\ref{nzerom}) of induced charged instanton zero modes are $n_1^+ = 1, \, n_1^-=0$ and 
 $n_2^+ = 0, \, n_2^-=1$, corresponding to precisely one zero mode $\eta_1$ and $\tilde \eta_2$. In this case
 the interaction term $\eta_1 \, \Phi  \, \tilde \eta_2$ allows for a saturation of the zero modes, thereby inducing said operator.
 However, all possible supersymmetric flux configurations have to be summed up in the instanton partition function.
A different flux $F_E$ giving rise to extra vectorlike pairs of zero modes, say $\chi_1,\tilde\chi_1$ on $\gamma_1$, could still be compatible with the same induced operator provided these are absorbed in the path integral by a higher order contact term in the instanton effective action, e.g. of the form   $\eta_1 \, \Phi  \, \tilde \eta_2 \, \chi_1 \, \tilde \chi_1$. Such choice of $F_E$ would correspond also to a different supersymmetric configuration of $T_3$-flux whose dual 3-chain nonetheless has the same boundary given by one copy of $\mathbb P^1_E$.


There are many possible variations of what we have just described. For instance, on top of the generation of couplings charged under massive $U(1)$s such as (\ref{chargeW}), the instanton can in principle contribute a full tower of uncharged localised operators.
For the above M5-instanton producing the $\Phi \, e^{- S_{M5}}$  term in the massive $U(1)$ model this is because 
the \emph{same} $T_3$ configuration is also compatible with $n$ copies of  M2-branes wrapping the fibral curve $\mathbb P^1_{E,1}$ and $n-1$ copies of M2-branes wrapping the orientation reversed $-\mathbb P^1_{E,1}$ over the \emph{same} point $p_1$ in the base. Indeed, it is still the case that $\partial \Gamma_{\rm np} = n \mathbb P^1_{E,1} - (n-1) \mathbb P^1_{E,1}$. This produces a full tower of couplings  $\sum_n \, \Phi^n \tilde \Phi^{n-1} e^{- S_{M5}}$, where the higher terms are suitably mass-suppressed. 
In Type IIB language the appearance of this tower of couplings corresponds to an instanton flux $F_E$ with the same number of zero modes as before such that  the instanton effective action contains not only the coupling $\eta_1 \Phi \tilde \eta_2$ (or its higher order analogue involving vectorlike pairs of zero modes), but likewise their higher order cousins $\sum_n \eta_1 \Phi^n \tilde \Phi^{n-1} \tilde \eta_2$. Integrating out the charged instanton zero modes results in the corresponding operators. 

Alternatively, one can consider an M5-instanton with 3-form flux $T_3$ dual to a different chain $\Gamma_{\rm np}$ bounded by $n$ copies of $\mathbb P^1_E$ and $n-1$ copies of $-\mathbb P^1_E$ over mutually \emph{different} points in the base. 
The $n-1$ pairs of M2-branes on distinct copies of $\mathbb P^1_E $ and $-\mathbb P^1_E $ wrap perturbatively inequivalent fibral curves on $D$, see equ. (\ref{P1kP1l}). They do therefore not cancel locally, and compensating $T_3$ flux must be introduced to satisfy the Bianchi identity. These couplings can in principle be distinguished quantitatively from the ones discussed above because 
the wavefunction associated with the physical modes $\Phi$ and $\tilde \Phi$ varies over the matter curves; contributions from different points $p_k$ therefore lead to different answers, and all consistent configurations must be summed up to obtain the instanton partition function.

\subsection{Non-perturbative Yukawa hierarchies from fluxed M5-instantons}

We now describe how fluxed  M5-instantons of the kind discussed in this section  can generate contributions to the Yukawa couplings in an F-theory compactification. 
Apart from serving as an interesting application, such non-perturbative couplings can be of considerable phenomenological interest
in GUT model building. To compute the final expression for Yukawa couplings one must sum up, as always, both the perturbative, i.e.\ tree-level, contribution  as well as all non-perturbative corrections. The M5-instanton effect to be described momentarily is a qualitatively new type of such corrections which has not been considered in F-theory before. It will in general modify the hitherto obtained results e.g. for the rank of the Yukawa couplings. 
In particular the M5-instanton effect in question represents a different type of non-perturbative correction compared to the types of instanton contributions introduced in \cite{Marchesano:2009rz}, which also modify the tree-level expression for the Yukawa couplings in F-theory model building \cite{Aparicio:2011jx,Font:2012wq,Font:2013ida,Marchesano:2015dfa}.

In order to illustrate the general idea, we consider for concreteness the $SU(5)$ GUT model reviewed in section 4 of \cite{Martucci:2015dxa} (and studied previously in the references therein). 
In the base there are two matter curves, $\calc_{\bf 5}$ and $\calc_{\bf 10}$, supporting matter transforming in the ${\bf 5}$ and ${\bf 10}$ representation respectively under the $SU(5)$ gauge group. 
In the notation of \cite{Martucci:2015dxa}, on $\calc_{\bf 5}$ one can explicitly identify two fibral curves $\mathbb P^1_{3D}$ and $\mathbb P^1_{3C}$ representing states in the ${\bf 5}$ and ${\bf \bar 5}$ representation respectively, while on  $\calc_{\bf 10}$ there are fibral curves $\mathbb{P}^1_{32}$, $\mathbb{P}^1_{4D}$ and $\mathbb{P}^1_{24}$ representing states in the ${\bf 10}$, ${\bf 10}$ and ${\bf \overline{10}}$ representations respectively. 

The matter curves $\calc_{\bf 5}$ and $\calc_{\bf 10}$  generically intersect at two point sets $p_{E_6}$ and $p_{D_6}$ on the base $B$.  By moving the above fibral curves along the matter curves, they can merge by a splitting and joining process at $p_{E_6}$ and $p_{D_6}$. This results, respectively,  in the following perturbative homological relations
\be\label{bulkSU(5)}
[C_{\bf 10 \, 10 \, 5}]_{\rm p}=0   \quad,\quad [C_{\bf 10 \, \bar 5 \,  \bar 5}]_{\rm p}=0  \,,
\ee
with
\be
C_{\bf 10 \, 10 \, 5}\equiv   \mathbb{P}^1_{24}+\mathbb P^1_{3C} -\mathbb{P}^1_{4D}     \quad,\quad    C_{\bf 10 \, \bar 5 \,  \bar 5}\equiv  \mathbb{P}^1_{32}+\mathbb{P}^1_{3C}-\mathbb P^1_{3D} \,.
\ee
Such bulk perturbative relations allow for the generation, at the perturbative level, of the couplings ${\bf 10 \, 10 \, 5}$ and ${\bf 10 \, \bar 5 \,  \bar 5}$ respectively, in the four-dimensional effective action. 

Consider now an M5-instanton along a divisor $D = \pi^{-1} D_{\rm b}$. Generically, it intersects the matter curves $\calc_{\bf 5}$ and $\calc_{\bf 10}$ at  isolated point sets $p_{\bf 5}$ and $p_{\bf 10}$ respectively. Hence the above fibral curves in the bulk restrict to fibral curves on $D$. However, the fibral curves are now stuck at the points $p_{\bf 5}$ and $p_{\bf 10}$ and, since generically  they differ from $p_{E_6}$ and $p_{D_6}$, on the M5-brane we have $[C_{\bf 10 \, 10 \, 5}]_{D,{\rm p}}\neq 0$ and  $[C_{\bf 10 \, \bar 5 \,  \bar 5}]_{D,{\rm p}}\neq 0$.
On the other hand, as discussed in section \ref{sec_pertvsnonpert}, homological triviality in the bulk generically implies non-perturbative homological triviality on $D$. Hence, we generically have
\be
 [C_{\bf 10 \, 10 \, 5}]_{D,{\rm np}}= 0\quad,\quad [C_{\bf 10 \, \bar 5 \,  \bar 5}]_{D,{\rm np}}= 0  \,.
\ee
This implies that one can turn on a non-closed $T_3$ flux such that either $\d T_3 = - \delta_{4}( C_{\bf 10 \, 10 \, 5}) $ or $\d T_3 = - \delta_{4}( C_{\bf 10 \, \bar 5 \,  \bar 5}) $. The corresponding fluxed M5-brane instantons, if supersymmetric and with the proper number of fermionic zero modes, can then generate corrections to the superpotential of the form
\be
\Phi_{\bf 10} \,  \Phi_{\bf 10} \,  \Phi_{\bf 5} \, e^{-S_{M5}}\quad,\quad \Phi_{\bf 10} \,  \Phi_{\bf \bar 5} \,  \Phi_{\bf \bar 5} \, e^{-S_{M5}}.
\ee
These have to be added to the perturbatively generated Yukawa couplings, which have rank one,  basically because their generation 
can be described ultra-locally around the points $p_{E_6}$ and $p_{D_6}$ \cite{Cecotti:2010bp}. The above contribution from fluxed M5-brane instantons is instead delocalised and is expected to generically violate the rank one property. Hence, this may provide a natural mechanism, different from the mechanism proposed in  \cite{Marchesano:2009rz}, for generating  hierarchical Yukawa couplings.
It would be very interesting to develop computational tools to more explicitly evaluate such corrections to the Yukawa couplings.

The appearance of such a fluxed M5-instanton contribution is in agreement with expectations with analogous weakly coupled Type IIB models.
If the perturbative Yukawa points $p_{E_6}$ at which the  ${\bf 10 \, 10 \, 5}$ coupling is localised in F-theory is absent, a smooth  Sen limit can be taken. 
An example for an F-theory $SU(5)$ Tate model with such a Sen limit has been given in \cite{Blumenhagen:2009up}, where the points $p_{E_6}$ do not arise due to the specific intersection numbers of the base space. 
In weakly coupled type IIB models, the $SU(5)$ gauge group is known to be accompanied by a geometrically massive $U(1)$ symmetry from the diagonal $U(1)$ in $U(5)$.
This massive $U(1)$ symmetry obstructs a coupling ${\bf 10}_2 \, {\bf 10}_2 \, {\bf 5}_1$ at the perturbative level, where the subscripts denote the $U(1)$ charges. 
As reviewed in appendix \ref{chargedD3inst}, even in absence of gauge flux on the 7-branes  a D3-instanton can compensate for this $U(1)$ charge if it carries orientifold-odd instanton flux $F_E$. This is precisely the dual configuration to the fluxed M5-instanton considered above. 
We will give more details on the Type IIB construction of such fluxed instantons in appendix \ref{Appendix-IIBExample}, albeit for an instanton with non-trivial pullback of the 7-brane gauge flux.

\section{M5 selection rules: second case  ($G_4|_D\neq 0$ and $T_3=0$)}
\label{sec:T3=0}

Let us now suppose that $G_4|_D\neq 0$ or, more precisely, that the dual algebraic cycle $C_G$ introduced in section \ref{sec_pertvsnonpert}  is non-vanishing. 
We can then write (\ref{T3BI}) in the alternative form
\be\label{BIref}
\d T_3=\delta_4(C_G)-\delta_4(C),
\ee
which should be understood up to rational equivalence of $C_G$ and perturbative homological equivalence of  $C$.\footnote{Notice that changing the rational equivalence class of the algebraic 4-cycle $\cald_G$ associated with a certain $G_4$ flux, but not its homology class, corresponds to a change of the `Wilson line' associated with the potential $C_3$. Such change restricts to a change of the rational equivalence class of $C_G$ within the same homology class in $D$ and, according to (\ref{BIref}), must be compensated by a corresponding change of $T_3$, consistently with the local split (\ref{localsplit}).}
 At the coarsest level, (\ref{BIref}) can be satisfied only if the tadpole condition 
\be
[C_G-C]_D=0
\ee
is fulfilled. At a more refined level we can either have $[C_G-C]_{D,{\rm p}}=0$ or just $[C_G-C]_{D,{\rm np}}=0$.

In this section we consider the simplest way to solve  (\ref{BIref}), which is by just inserting a number of time-like M2-branes wrapping an algebraic cycle $C$ perturbatively equivalent to $C_G=\sum_A n^A_G\,C_A$ so that
\be
[C_G-C]_{D,{\rm p}}=0.
\ee
In this case we can set $T_3=0$ and the M5-instanton produces an F-term of the form
\be\label{Gcharge}
\prod_A (\Phi_A)^{n_G^A} \, e^{ - S_{\rm M5}}
\ee
where, if $n^A_G<0$, $\Phi_A^{n^A_G}=\tilde\Phi_A^{|n^A_G|}$, with $\tilde\Phi_A$ being chiral field conjugated to $\Phi_A$. This situation corresponds, at weak coupling, to a D3-brane instanton intersecting two intersecting D7-branes whose world-volume flux generates non-trivial fermionic zero-modes. Their number is reflected in the number of inserted operators $\Phi_A$.

In models with geometrically massless $U(1)$s, the insertion of M2-branes  ensures that the $G_4$-induced charge of the M5-brane instanton with respect to all massless $U(1)$ gauge symmetries is cancelled by the charged operator insertions.\footnote{If the flux $G_4$ is chosen such as not to break the non-abelian gauge symmetries, i.e. $\int_{\hat \Y} [G_4] \wedge [E_i] \wedge [\pi^{-1} D_a] =0$ for all $[D_a] \in H^{1,1}(\B)$ and $E_i$ resolution divisors, the M5-instanton is automatically invariant under the Cartan $U(1)$s. Otherwise the gauge group is broken and analogous statements can be made with respect to the new abelian gauge factors which arise from the Cartan subalgebra of the original gauge group.} However, (\ref{Gcharge}) contains more refined information about the operator that needs to be inserted. This can be regarded  as a manifestation of hidden massive $U(1)$ symmetries which induce additional selection rules.

\subsection{An example in an $SU(5)$ model} 
\label{sec:fluxSU(5)}

We now discuss a realisation  of the type of non-perturbatively correction  considered in this section  in an explicit model. 
Our example will clearly illustrate the difference between situations in which $[G_4|_D] \neq 0 \in H^4(D)$ and thus an insertion of M2-brane states is obviously required to solve the Bianchi identity, as opposed to the more subtle case in which $[G_4|_D] = 0 \in H^4(D)$, but $[C_G] \neq 0 \in {\rm CH}^2(D)$. From the above discussion we know that also such configurations require the introduction of M2-branes or 3-form flux to saturate the Bianchi identity, and we will see how this comes about.

We begin by exemplifying the first type of  situation by means of a $U(1)$ restricted Tate model with gauge group $SU(5) \times U(1)_X$ as first described in \cite{Krause:2011xj}. The non-abelian gauge group arises from 7-branes wrapping an effective divisor $W$ defined by the equation $w=0$, where $w$ is a section of a line bundle $\call_W$ over the base space $\B$. The resolved elliptic fibration $\hat Y$ is described by the vanishing of the hypersurface constraint 
\bea \label{Su5U1X}
\hat P_T& :=& e_1 \, e_2^2 \, e_3 \, s^2 \, x^3 - e_3 \, e_4 \, s \, y^2 - a_1 \, x \, y \, z \, s + a_{21} \, e_0 \, e_1 \, e_2 \, x^2 \, z^2 \, s  - 
   a_{32} \, e_0^2 \, e_1 \, e_4 \, y \, z^3  \\
&&+ a_{43} \, e_0^3 \, e_1^2 \, e_2 \, e_4 \, x \, z^4 \nonumber 
   \eea
   within an ambient 5-fold. 
On $\hat Y$ the resolution divisors $E_i: \{e_i=0\}$ are rationally fibered over the divisor $W$ in the base with fibre $\mathbb P^1_i$. Their dual cohomology classes $[E_i]$ generate the Cartan $U(1)$s of $SU(5)$.
The coefficient $a_1$  transforms as a section of the anti-canonical bundle $\bar K$ of the base, and similarly 
$a_{nk}$ are sections of the line bundles $\bar K^n\otimes \bar\call^k_W$. The $U(1)_X$ symmetry is associated with the existence of a (1,1)-form ${\rm w}_X$ defined at cohomological level by
\be\label{wXdef}
[{\rm w}_X]= [5(S-Z-\bar K)+2E_1+4 E_2+6 E_3+3E_4] \ \in\  H^{1,1}(\hat Y).
\ee
Here and in the following Poincar\'e duality and the identification of line bundles with the corresponding (equivalence classes of) divisors are understood.

\begin{figure}[t]
    \centering
  \begin{tikzpicture}[scale=3.9,main node/.style={circle,
  draw,minimum size=4mm}]
  \node[main node] (1) {$3C$};
  \node[main node] (2) [above left of=1] {$24$};
  \node[main node] (7) [above right of=2] {$2B$};
  \node[main node] (4) [left of=2] {$14$};
  \node[main node] (5) [above left of=4] {$0A$};
  \node[main node] (6) [below left of=4] {$4D$};
    \draw[ultra thick] (1) -- (2);
      \draw[ultra thick] (2) -- (7);
    \draw[ultra thick] (2) -- (4);
       \draw[ultra thick] (4) -- (5);
    \draw[ultra thick] (4) -- (6);
  \end{tikzpicture}
      \caption{Fibre   topology over the curve $C_{\bf 10}$, corresponding to triangulation $T_{11}$ of \cite{Krause:2011xj}.  The fibral curves arise from the fibral curves $\mathbb P^1_i$ corresponding to the simple roots over $C_{\bf 10}$ via the splittings $\mathbb P^1_1 \rightarrow \mathbb P^1_{14}$,  $\mathbb P^1_2 \rightarrow \mathbb P^1_{2B} +\mathbb P^1_{24} $,   $\mathbb P^1_3 \rightarrow \mathbb P^1_{3C} $, $\mathbb P^1_4 \rightarrow \mathbb P^1_{4D} +\mathbb P^1_{14}  + \mathbb P^1_{24}$.  \label{FiberSO10}}
\end{figure}
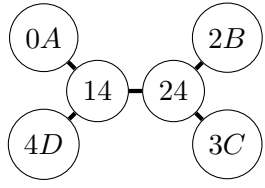

More information on the model and the notation can be found in  \cite{Krause:2011xj}.
In particular, we will need the fact that over the curve $\calc_{\bf 10} =  \{w=0\} \cap \{a_1 = 0\}$  the fibre takes the form of an affine Dynkin diagram of the Lie algebra ${\mathfrak{so}}(10)$, see figure \ref{FiberSO10}. M2-branes wrapping suitable linear combinations of fibral curves over $\calc_{\bf 10}$ give rise to matter states in the ${\bf 10}$ and ${\overline{\bf 10}}$-representation of $SU(5)$.

A possible `vertical' gauge flux,  first described in \cite{Marsano:2011hv} in the pure $SU(5)$ model without $U(1)$ restriction, is given at the cohomological level by
\be \label{G4fluxa}
[G_4] = 5 [\cals_{24}] + [\pi^{-1} \bar K] \wedge [2E_1 - E_2 + E_3 - 2 E_4] \in H^{2,2}({\hat \Y}).
\ee 
 Here $\cals_{24}$ denotes  the holomorphic 4-cycle given as the complete intersection 
\be
\cals_{24} = \{e_2=0\} \cap \{e_4=0\} \cap \{ a_1 =0 \}
\ee
 within the ambient space of $\hat \Y$, which lies automatically on the hypersurface $\hat \Y$. 
The surface $\cals_{24}$  has the structure of a rational fibration over the curve $\calc_{\bf 10} $ in the base. 
The correction terms $[\pi^{-1} \bar K] \wedge [2E_1 - E_2 + E_3 - 2 E_4]$ in (\ref{G4fluxa}) are introduced to ensure that $G_4$ does not break $SU(5)$ gauge invariance.

An M2-brane wrapping the rational fibre $\mathbb P^1_{24}$ of $\cals_{24}$ gives rise to a state with $SU(5)$ weight $-\mu_{10} + \alpha_1 + \alpha_2 + \alpha_3$.\footnote{ See \cite{Krause:2011xj} for details, especially table A.19; we are referring here to the specific triangulation $T_{11}$.} 
In our conventions an M2-brane wrapping $\mathbb P^1_i$ corresponds to a simple root $-\alpha_i$. The explicit weight vectors and simple roots are given by  $\mu_{\bf 10} = [0,1,0,0]$ and  $ \alpha_1 = [2,-1,0,0]$,    $ \alpha_2 = [-1,2,-1,0]$,   $ \alpha_3= [0,-1,2,-1]$,   $ \alpha_4 = [0,0,-1,2]$.



In order to apply the formalism described above we need to define not only the  class of $G_4$ as an element of $H^{2,2}(\hat \Y)$, but to specify a rational equivalence class of dual algebraic 4-cycles. Since the notion of (co)homology is coarser than the notion of rational equivalence, this choice of an element in ${\rm CH}^2(\hat Y)$ need in general not be unique.
For ease of notation we will use the same symbol to denote an algebraic cycle and its rational equivalence class. 
The first term in (\ref{G4fluxa}) is manifestly given by the cohomology class of the algebraic 4-cycle $\cals_{24}$. As for the correction terms, the cohomology class $[\bar K]$ on the base defines a unique divisor class (up to rational equivalence) since $B$ is simply connected (and we assume the absence of torsion  for simplicity). 
A simple representative is provided by the divisor $D_{a_1}$ described by the vanishing of the Tate polynomial $\{a_1 = 0\}$ on the base.
Hence we can take the equivalence class of algebraic 4-cycles
\be \label{Dg1}
{\cal D}_G = 5 \,  \cals_{24} + \pi^{-1} D_{a_1} \cdot (2 E_1 - E_2 + E_3 - 2 E_4)\ \in \ {\rm CH}^2(\hat Y)\,
\ee
as the definition of our $G_4$ flux.
For a review of the notion of intersection product used in this definition we refer e.g. to the discussion in section 2.4 of \cite{Bies:2014sra}.

The next step is to compute the element $C_G \in {\rm CH}^2(D)$ obtained by restricting ${\cal D}_G$ to the M5 divisor $D=\pi^{-1}D_{\rm b}$.
The restriction of $\cals_{24}$ to $D$ is given by a copy of $\mathbb P^1_{24}$ in the fibre over each of the points obtained by intersecting  $\calc_{\bf 10}$ with $D_{\rm b}$.
Let us assume that $\calc_{\bf 10}$ and $D_{\rm b}$ intersect in $n$ effective points $p_l$. Then we can write this as
\be
\cals_{24} |_D = \sum_{l=1}^n \big(\mathbb P^1_{24}\big)_l  \in {\rm CH}^2(D)  ,
\ee
where $\big(\mathbb P^1_{24}\big)_l$ denotes said fibral curve over the point $p_l$. 

As for the remaining terms in (\ref{Dg1}), each resolution divisor $E_i$ intersects $D$ in a rational surface given by fibering $\mathbb P^1_i$ over the curve $W \cap D_{\rm b}$.
The intersection of this curve with $D_{a_1}$ is given by the \emph{same} points $p_l$ because $D_{a_1}$ has been chosen such that $D_{a_1} \cap W = C_{\bf 10}$.\footnote{For a more general divisor $\tilde D_{a_1}$ in the same divisor class, the pullback $\pi^{-1} \tilde D_{a_1} \cdot (2 E_1 - E_2 + E_3 - 2 E_4)|_{D}$  is rationally equivalent to $\pi^{-1}  D_{a_1} \cdot (2 E_1 - E_2 + E_3 - 2 E_4)|_{D}$ on $D$ - see footnote \ref{footnoteChow1}. This matches the fact that we can holomorphically transport the fibral curves $\mathbb P^1_i$ over the curve $W \cap D_{\rm b}$ from the new intersection points $\tilde p_l = \tilde D_{a_1} \cap W \cap D_{\rm b}$ to the intersection points $p_l = D_{a_1} \cap W \cap D_{\rm b}$. Hence any choice of divisor in class $\bar K$ induces the same instanton coupling.}
Altogether therefore 
\bea  \label{pullback1}
C_G = \sum_{l=1}^n   C_{G,l}  \in {\rm CH}^2(D), \qquad C_{G,l} = \big(5 \,  \mathbb P^1_{24} + 2 \,  \mathbb P^1_1 - \mathbb P^1_2+  \mathbb P^1_3 - 2 \, \mathbb P^1_4\big)_l  \, , 
\eea
where the object $C_{G,l}$ is (the rational equivalence class of) the indicated fibral curves in the fibre ${\mathfrak f}|_{p_l}$ of $D$.
Notice that the homology class $[C_{G}]_D \in H_2(D)$ is non-vanishing and $[G_4|_D] \neq 0 \in H^4(D)$ as advertised. This is because the unique intersection point of the rational section 
$S:\{s=0\}$  with the fibre lies on $\mathbb P^1_3$ and thus $\int_{\mathbb P^1_3}  S = 1$. In fact, $S$ is the only divisor on $\hat Y$ with a net intersection number with $C_{G}$, and this holds by pullback for all divisors on $D$, provided $D$ is chosen sufficiently generic.

One can then locally cancel the term $\delta_4(C_G)$ in the Bianchi identity by a configuration of M2-branes wrapping the  curves $C_{G,l}$ over each of the $n$ points $p_l$. 
Since this cancellation is local, at each point $p_l$,  this is possible for $T_3 \equiv 0$. 
 Indeed,  $C_{G,l}$ is \emph{perturbatively}  equivalent\footnote{To see this use the fact that the splitting of curves $\mathbb P^1_i$ over $C_{\bf 10}$ implies that $[\mathbb P^1_4]_{\rm p} = [\mathbb P^1_{4D} +\mathbb P^1_{14}  + \mathbb P^1_{24}  ]_{\rm p} $,  $[\mathbb P^1_2]_{\rm p} = [\mathbb P^1_{2B} +\mathbb P^1_{24} ]_{\rm p}$,   $[\mathbb P^1_3]_{\rm p} = [\mathbb P^1_{3C} ]_{\rm p}$, $[\mathbb P^1_1]_{\rm p} = [\mathbb P^1_{14}]_{\rm p}$, see figure \ref{FiberSO10}. These relations also hold as perturbative homological relations on $D$ for generic $D$. } e.g.\ to the combination of fibral curves
\be\label{decfibral}
[C_{G,l}]_{D,{\rm p}} = \Big[   \big(\mathbb P^1_{3C} + \mathbb P^1_{24})_l  + \big(\mathbb P^1_{14} + \mathbb P^1_{24}\big)_l + \big(-\mathbb P^1_{4D}\big)_l + (-\mathbb P^1_{2B})_l + \big(-\mathbb P^1_{14} - \mathbb P^1_{4D}\big)_l\Big]_{D, {\rm p}}.
\ee
This corresponds to the decomposition of the state associated with an M2-brane wrapping $C_{G,l}$ into  the following sum of five ${\bf \overline{10}}$ weights 
\be\label{weightdec}
\begin{aligned}
&(- \mu_{10} + \alpha_1 + \alpha_2) +  (- \mu_{10} + \alpha_2 + \alpha_3) +  (- \mu_{10} + \alpha_2 + \alpha_3 + \alpha_4)
+  (- \mu_{10} + \alpha_1 + 2 \alpha_2 + \alpha_3) +   \\
&  (- \mu_{10} + \alpha_1 + \alpha_2 + \alpha_3 + \alpha_4)=[0,0,0,0],
\end{aligned}
\ee
see table (A.19) in \cite{Krause:2011xj}.
Each of the five weight vectors appearing on the l.h.s.\ of (\ref{weightdec}) gives the Cartan charges of each of the five fibral curves appearing on the r.h.s.\ of (\ref{decfibral}), 
 over any of the $n$ intersection points $p_l$. They sum up to a state of zero Cartan charges basically because the flux $G_4$ has been chosen such as not to break the $SU(5)$ gauge symmetry. Thus $C_{G,l}$ can be identified with the rational equivalence class of the sum of these five fibral curves. An M2 wrapping a representative in this class gives rise to an $SU(5)$ singlet obtained from five copies of states in the ${\overline{\bf 10}}$ representation,
\be\label{statesing}
C_{G,l}\sim ({\overline{\bf 10}} \, {\overline{\bf 10}} \,  {\overline{\bf 10}} \, {\overline{\bf 10}} \, {\overline{\bf 10}})_{\rm sing},
\ee
where gauge invariant contraction of the $SU(5)$ indices is understood.\footnote{In fact, there are $6$ inequivalent gauge invariant contractions \cite{Feger:2012bs}. To determine which of these are generated a more detailed analysis is required, which are not performing here.}

If we solve the Bianchi identity by a configuration with precisely these M2-branes, the induced operator is 
\be \label{105coupling}
({\overline{\bf 10}} \, {\overline{\bf 10}} \,  {\overline{\bf 10}} \, {\overline{\bf 10}} \, {\overline{\bf 10}})_{\rm sing} ^n  \, e^{- S_{\rm M5}}.
\ee
The state singlet (\ref{statesing}) has charge $+5$ under the $U(1)_X$ gauge symmetry as can be readily checked by computing the intersection number  $\int_{C_{G,l}} \tw_X=5$ from (\ref{wXdef}) and (\ref{pullback1}). Correspondingly, the M5-brane instanton action is anomalous in such a way that $e^{-S_{M5}}$ carries a $U(1)_X$ charge of $-5n$.
Indeed, under a gauge transformation (\ref{C3gauge}) with $\Lambda_2 = \lambda \,  \tw_X$ and $\lambda$ the  gauge parameter in the external dimensions one finds
\be
e^{-S_{M5}} \rightarrow     e^{-2 \pi\ii \lambda \sum_l \int_{C_{G,l}} \tw_X}    e^{-S_{M5}} = e^{- 2 \pi \ii (5 n \lambda) }  e^{-S_{M5}}.
\ee
Hence, consistently, the operator (\ref{105coupling}) is invariant under the $U(1)_X$ gauge symmetry as well.


Having spelled out in some detail the case with gauge group $SU(5) \times U(1)_X$, we can immediately extend our considerations to a generic $SU(5)$ Tate model. This can be obtained by setting $s=1$ in (\ref{Su5U1X})  and adding a term $a_{65} \, e^5_0 \,e^3_1 \, e_2 \,  e^2_4 \, z^6$. In this case the rational section $S$ is absent and along with it the (1,1)-form (\ref{wXdef}). Hence the $U(1)_X$  gauge field cannot be defined nor can the notion of a $U(1)_X$ charge of the states. Apart from this, the resolved fibre over the ${\bf 10}$-curve does not change significantly, and also the flux (\ref{G4fluxa}) remains a valid gauge flux. The crucial difference is that now, its pullback 
$G_4|_D$ vanishes in $H^4(D)$. As stressed after (\ref{pullback1}), in the $U(1)$ restricted model the extra rational section $S$ is the only divisor with non-trivial intersection with $C_G$. In its absence $C_G$ has vanishing intersection number with all divisors on $\hat Y$ and, generically, also with all divisors on $D$. Therefore $[C_G]_D=0 \in H_2(D)$. Nonetheless, $[C_G]_{D,{\rm p}} \neq 0$ (so that $C_G \neq 0 \in {\rm CH}^2(D)$ ) and then we can only write $[C_G]_{D,{\rm np}} = 0$. To see this note first that no local relation in the homology of each isolated fibre ${\mathfrak f}_l$ over $p_l$ exists according to which $[C_G]_{D,{\rm p}} = 0$. The only possibility would then be that $[C_G]_{{\rm p}} = 0$ in the bulk,  due to an interpolating perturbative 3-chain  with one leg on  ${\cal C}_{\bf 10}$. Since a generic instanton divisor $D_{\rm b}$ intersects  ${\cal C}_{\bf 10}$ in isolated points, this 3-chain would restrict to a non-perturbative one on $D$.

The Bianchi identity can still be solved by insertion of M2-branes as before, which gives rise to the same coupling (\ref{105coupling}). This time, however, considerations based on cohomology alone, or equivalenty on the $U(1)$ charges of the states and the M5-brane, would not be sufficient to deduce the form of the coupling. 

In the generic $SU(5)$ Tate model, another possibility to solve the Bianchi identity (\ref{BIref}) would be to turn on a $T_3$ flux dual to the 3-chain $\Gamma$ trivialising   $C_G$ within $D$, as  discussed in the next section.

\section{M5 selection rules: third case  ($G_4|_D\neq 0$ and $T_3\neq 0$)}
\label{sec_thirdcase}

Finally we consider the possibility of a combined presence of a background $G_4$ flux and of a non-vanishing world-volume $T_3$ which partially or completely cancels the $G_4$ contribution to the Bianchi identity (\ref{BIref}), assuming that the latter corresponds to an algebraic cycle $C_G$ which is non-vanishing in the rational homology of $D$. 
This situation occurs if the time-like M2-branes inserted in the path-integral wrap a fibral cycle $C\subset D$ which is homologous to $C_G$ in $H_2(D)$ (that is, (\ref{cohoBI}) is indeed satisfied)  but not in the sense of perturbative homological equivalence on $D$: $[C_G-C]_{D,{\rm p}}\neq 0$. In other words, 
\be\label{npGC}
[C_G-C]_{D,{\rm np}}=0,
\ee
and there exists a non-trivial non-perturbative 3-chain $\Gamma_{\rm np}$ with non-vanishing volume in the F-theory limit and such that
\be
\del\Gamma_{\rm np}=C_G-C\,.
\ee 
Consequently we must turn on a non-closed $T_3$ that can be regarded as the Poincar\'e dual of $\Gamma_{\rm np}$, 
as we did in section \ref{sec:G4trivial}. By expanding $C=\sum_A n^A C_A$
into fibral curves, such M5-instanton produces an F-term proportional to
\be\label{insertions2}
\prod_A (\Phi_A)^{n^A} \, e^{ - S_{\rm M5}},
\ee
where for $n^A<0$ we mean $\Phi_A^{n^A}\equiv \tilde\Phi^{|n^A|}_A$. The prefactor $\prod_A (\Phi_A)^{n^A}$ is well-defined only up to possible perturbatively allowed operators, which amount to changing $C$ by 2-cycles that are perturbatively trivial.   

This situation includes the case in which, at weak coupling, there are both bulk D7-brane fluxes $F_a$ and D3-brane instanton flux $F_E$, such that the D3-instanton charges (\ref{E3charges}) receive contributions from both fluxes. Notice that (\ref{npGC}) implies that the M2-brane states in fact detect some hidden massive $U(1)$ gauge symmetry, which is associated with a corresponding selection rule. This is the M/F-theory counterpart of the observation stressed in appendix \ref{chargedD3inst} that, at weak coupling, the  flux $F_E$ can contribute to the D3-instanton charges (\ref{E3charges})  only in presence of geometrically massive $U(1)$s.
In appendix \ref{Appendix-IIBExample} we exemplify how this interplay of $F_E$ and $F_a$ can generate non-perturbative Yukawa couplings in a Type IIB setting with a geometrically massive $U(1)$.

As a particular subcase, suppose that $C_G$ is trivial in ordinary homology, but $[C_G]_{D,{\rm p}}\neq 0$ so that
 \be\label{npG}
[C_G]_{D,{\rm np}}=0
\ee
as for instance in the setting considered at the end of section \ref{sec:fluxSU(5)}.
In this situation it is consistent to set $C=0$ so that the M5-brane instanton can contribute to an F-term  which is not dressed by any matter operator due to the cancellation of $C_G$ by instanton flux,
\be
\d T_3 = \delta_4(C_G).
\ee
 This is an important effect in the context of moduli stabilisation which had been anticipated in \cite{Grimm:2011dj} for weakly coupled Type IIB orientifold compactifications, where the D3-instanton flux cancels the gauge flux induced charge of the instanton with respect to a geometrically massive $U(1)$.

In the remainder of this section we elucidate how this can be realised in the language of M5-instantons. Since massive $U(1)$ gauge flux is notoriously difficult to describe explicitly in global F-theory models (see however \cite{Krause:2012yh}), we will exploit the stable degeneration limit \cite{Clingher:2012rg,Donagi:2012ts} of the massive $U(1)$ model of \cite{Braun:2014nva}. Along the way, we will gain further insights into the structure of massive $U(1)$ gauge flux which are interesting also independently of the applications to the M5-instantons.


\subsection{Digression: Massive $U(1)$ gauge flux in the stable degeneration}

The Sen limit \cite{Sen:1996vd,Sen:1997gv} of an F-theory model can be described in terms of a complex family of elliptically fibered 4-folds $Y_\epsilon$ with the limit $\epsilon \rightarrow 0$ parametrising the weak coupling regime. One can describe such limit as a stable degeneration \cite{Clingher:2012rg,Donagi:2012ts} in which the 4-fold $Y_\epsilon$ degenerates to
\be
\Y_0=Y_E\cup_X Y_T.
\ee 
Here $Y_E$ is a $\mathbb{P}^1$-fibration over the base $\B$. Its degeneration locus $\Delta_E=0$ defines the position of the D7-branes in the weak-coupling limit. 
$Y_T$  is also a $\mathbb{P}^1$-fibration over $\B$, but it never degenerates. In particular, it has a global section, with associated divisor $Z$, inherited from the degenerated elliptically fibered 4-fold. Furthermore $Y_E$ and $Y_T$ intersect on the Calabi-Yau 3-fold $X$, which coincides with the IIB orientifold double cover of the base $\B$ with projection map $p: \X\rightarrow \B$.

As discussed in  \cite{Donagi:2012ts,Clingher:2012rg}, the cohomological structure of the nearby non-degenerated elliptic fibration $Y_\epsilon$ is encoded in the logarithmic de Rham cohomology groups $H^k_{\rm log}(Y_0)$, which can in principle be identified through the long exact sequence 
\be\label{logH}
\ldots\  \rightarrow \ H^{k-2}(X)\ \rightarrow \  H^{k}(Y_0)\  \rightarrow \  H^{k}_{\rm log}(Y_0)\  \rightarrow\   H^{k-1}(X)\  \rightarrow\  \ldots
\ee
The cohomology  of $Y_0$ in turn follows from the cohomology of $Y_T$ and $Y_E$ via the Meyer-Vietoris exact sequence. 
In particular, we will need the fact that elements of $H^k(Y_E)$ define non-trivial elements of $H^k(Y_0)$ only if their restriction to $X$ is trivial in $H^k(X)$ or even under the orientifold projection $\sigma: X \rightarrow X$. 
A rather detailed explanation of this is can be found in section 6 of \cite{Martucci:2015dxa}.


One can exploit these relations to identify the 2-cocyles on $Y_0$ which generate the uplift of the $U(1)$ symmetries of the IIB model.
The D7-branes are located at the discriminant locus $\Delta_E=0$, over which the $Y_E$ fibre degenerates. 
Suppose that $\Delta_E$ splits into into irreducible components  $\Delta_E=\prod_a \Delta_a^E$ so that $\{\Delta_a^E=0\}$ defines the locus of a single connected D7$_a$-brane. 
Over each component $\Delta_a$ the fibration locally splits  into two rational fibrations $R^+_a$ and $R^-_a$.
Correspondingly, for each D7$_a$-brane,  one can locally construct the divisor \cite{Braun:2014pva}
\be
S^E_a=\frac{1}{2}(R^+_a-R^-_a).
\ee
Notice that such divisor is not always globally well-defined,  for instance when the corresponding D7-brane wraps a single orientifold invariant divisor on the double cover Calabi-Yau $\X$.
By Poincar\'e duality, $S^E_a$ defines a 2-cocycle ${\rm w}^E_a\equiv \frac{1}{2} [R^+_a-R^-_a]\in H^2(Y_E)$ on $Y_E$, associated with the $U(1)$ gauge field supported on $\{\Delta_a^E=0\}$. 
As mentioned above,  ${\rm w}^E_a$ extends to an element of $H^2(Y_0)$ if it  restricts to an {\em even} or {\em trivial} class on $\X$. 
Now, the divisor $R^+_a$ intersects $X$ on the divisor $D_a\subset X$ wrapped by the D7$_a$ brane, while  $R^-_a$  intersects $X$ on its orientifold image  $D_a'=\sigma_*D_a$. 
Thus  ${\rm w}^E_a$ restricts on $X$ to the {\em odd} 2-cochain  $\frac{1}{2}[D_a-D'_a]\in H^2(X)$. Hence, it can define a massless $U(1)$ only if it is trivial in $H^2(X)$, that is if $[D_a-D'_a]=0$ .  This perfectly matches the weakly-coupled IIB picture. In particular, the geometric St\"uckelberg mechanism is at work when $[D_a-D'_a]\neq 0$, which corresponds to the case in which ${\rm w}^E_a$ cannot be extended to an element of $H^2(\Y_0)$ \cite{Martucci:2015dxa}.

Now consider any flux $F_a$ on $D_a$ and its orientifold image flux $F'_a=-\sigma^*F_a$ on $D'_a$.\footnote{In the case of orientifold invariant $D_a\equiv \sigma_*D_a'$, consistency requires $F_a\equiv F_a'=-\sigma^*F_a$.}  We can take the respective Poincar\'e dual 2-cycles $\Sigma_a$ and $\Sigma'_a$ in $X$, such that $\Sigma'_a=-\sigma_*\Sigma_a$.  $\Sigma'_a$ and $-\Sigma_a$ project to the same 2-cycle $\calc_a$  on the base $\B$, $\calc_a=p_*\Sigma_a=-p_*\Sigma'_a$. Let us denote by $A^\pm_a$ the restriction of $R^\pm_a$ over $\calc_a$. Hence, if  $\frac12(A^+_a-A^-_a)$ is a globally well defined algebraic 4-cycle,  we can use the prescription given in  \cite{Braun:2014pva} and  introduce a $G_4$ flux on $Y_E$ as 
\be
G^E_4=\frac12\sum_a[A^+_a-A^-_a]\,.
\ee
 Notice that, by construction, $A^\pm_a$ are such that $A^+_a\cap X=\Sigma_a$ and $A^-_a\cap X=-\Sigma'_a$. Hence $G^E_4$ restricts on $X$ to the algebraic 2-cycle
\be
G^E_4|_X=\frac12\sum_a[\Sigma_a+\Sigma_a'],
\ee
which, crucially, is {\em odd} under the orientifold involution.
By the argument given after (\ref{logH}),
this implies that $G^E_4$ defines a 4-cocycle on the entire $Y_0$ if and only if $G^E_4|_X$ is trivial in cohomology, i.e.
\be\label{D5tadpole}
\sum_a(\Sigma_a+\Sigma_a')=0 \quad\text{in $H_2(X)$}\,.
\ee
But this is exactly the well-known D5-charge tadpole condition for the D7-brane world volume fluxes! The importance of the D5-tadpole condition in constructing massive $U(1)$ fluxes had been stressed already in \cite{Grimm:2011tb}.
Hence, once the tadpole condition (\ref{D5tadpole}) is satisfied, $G^E_4$ extends to a globally defined 4-form flux $G_4$ on $Y_0$. Notice that this construction works without any assumption on the $U(1)$ gauge fields supported on the D7-branes, which can be either massless or not.

\subsection{Cancellation of massive $U(1)$ flux on M5-instantons in the stable degeneration limit}

We are now armed to use this description of a `massive $U(1)$' gauge flux in the stable degeneration in order to analyse the restriction of such flux to an M5-instanton.
To this end consider the massive $U(1)$ model introduced in \cite{Braun:2014nva} and further investigated in section 6 of \cite{Martucci:2015dxa}.
The full F-theory model is defined in terms of an elliptically fibered 4-fold $Y$ which  asymptotes,
in the stable degeneration limit, to $ Y_0 = Y_E \cup_X Y_T$ with
\begin{subequations}
\begin{align}
Y_T& :  \quad \,  y^2 -  s^2    ( {(a_1^2 + w \, a_{21})} \, z^2 + s \, \lambda)=0, \\
 Y_E& : \quad \,  {y}^2 - ((a_1^2 + w \, a_{21}) \, s^2 + 2 \, w \, \eta  \, s \, t + w \, \chi \, t^2) = 0.
\end{align}
\end{subequations}
Here $[y : s : z]$ parametrise the fibre ambient space $\mathbb P_{231}$, and $t$ and $\lambda$ are related to the small coupling parameter $\epsilon$ defining the Sen limit as $Y_0 = {\rm lim}_{\epsilon \rightarrow 0} Y_\epsilon$. Finally, $a_{21}$, $\eta$, $\chi$, $w$ are suitable base sections.
The discriminant of the rational fibration $Y_E$,
\bea
\Delta_E=      w  \, \, [w \,  \eta^2 -   (a_1^2 + w \, a_{21}) \,  \chi  ],
\eea
describes the 7-brane locus on the base $\B$ of $Y_E$. The brane on the divisor $W: \{w=0\}$ corresponds to a non-homologous brane-image pair on the Type IIB double cover $X$. Its relative $U(1)$ is massive by the geometric St\"uckelberg mechanism. The remaining 7-brane wraps an orientifold-invariant Whitney brane on $X$ and carries no gauge group. The intersection of both branes splits into two curves
\bea
{\cal C}_I = \{w=0\} \cap \{a_1=0\}, \qquad {\cal C}_{II} = \{w=0\} \cap \{\chi=0\}.
\eea
The curve $C_{II}$ represents the intersection locus of the two 7-branes away from the orientifold, which wraps the divisor $\{a_1^2 + w \, a_{21}=0\}$. Over $C_{II}$  the full 4-fold $Y$ acquires an $I_2$ fibre singularity, indicating the presence of localised matter charged only under a massive $U(1)$ group  \cite{Braun:2014nva}. 
On $Y_E$, the singularity descends to a conifold singularity  at $y=s=0$ in the fibre over ${\cal C}_{II}$. It can be removed by a small resolution of $Y_E$ into
\bea \label{hatWEa}
\hat Y_E:  \qquad   (y+ a_1 s) \,  \lambda_1 =  w \, \lambda_2, \qquad    (y- a_1 s) \,  \lambda_2 =   ( a_{21} \, s^2 + 2 \eta \, s \, t + \chi \, t^2) \, \lambda_1
\eea
with $[\lambda_1 : \lambda_2]$ homogeneous coordinates on the resolution $\mathbb P^1$ dubbed $\mathbb P^1_E$ in the sequel. In particular, at $y=s=w=\chi=0$, $\lambda_1$ and $\lambda_2$ are unconstrained in $\hat Y_E$, which indicates the presence of $\mathbb P^1_E$. 
The full 4-fold $Y$ does not admit a crepant small resolution, but could be resolved into a non-K\"ahler space $\hat Y$.
While  $\mathbb P^1_E$ is homologically non-trivial on $\hat Y_E$, it is trivial in the homology of such $\hat Y$ \cite{Braun:2014nva}.
In fact, it is trivialised already on $Y_0$ by a non-perturbative 3-chain $\Gamma_{\rm np}$ extending into $Y_T$ \cite{Martucci:2015dxa} such that
\bea \label{3chainGamma}
\partial \Gamma_{\rm np} = -  \mathbb P^1_E     \qquad {\rm in} \quad Y_0.
\eea
In the language of \cite{Martucci:2015dxa}, $[\mathbb P^1_E]_{\rm np}=0$, but $[\mathbb P^1_E]_{\rm p} \neq 0$ on $\hat Y$ (and $Y_0$). M2-branes wrapping $\mathbb P^1_E$ give rise to the expected matter states
 charged under the massive $U(1)$.\footnote{
The fibre structure over ${\cal C}_{I}$ is more complicated. This curve represents 
 the simultaneous intersection of the 7-branes on top of the O7-plane. Away from the stable degeneration, the fibre of $Y$ over ${\cal C}_{I}$ merely acquires a type $II$ cuspidal singularity so that $Y$ is smooth as a 4-fold. 
This agrees with the fact that the strings between both 7-branes located on top of the O7-plane are projected out and no massless matter resides here \cite{Braun:2014nva,Martucci:2015dxa}.
The degeneration of the fibre over ${\cal C}_I$ is thus to be viewed as an artifact of the stable degeneration.}

This massive $U(1)$ symmetry and its associated gauge flux can be understood with the help of the cylinder construction which was generally introduced in the context of the Sen limit in \cite{Clingher:2012rg} and further worked out in \cite{Braun:2014nva} to concretely describe the uplift of geometrically massless $U(1)$s and their fluxes. 
 To this end, note that $\hat Y_E|_W$ splits into $R^+ \cup R^-$ with
\be
\begin{aligned}
&R^-:  \{w=0\}\cap \{y-a_1 s = 0\}  \cap \{\lambda_1 = 0\}  , \\
&R^+:  \{w=0\}\cap \{y+a_1 s = 0\} \cap \{(s^2 \, a_{21} + 2 \, \eta \, s \, t + \chi \,  t^2) \, \lambda_1 + 2 \, s \, \lambda_2 \, a_1 =0 \} .
\end{aligned}
\ee
We can then apply the general procedure outlined above, construct a globally well defined divisor $\frac12(R^+-R^-)$  and introduce the Poincar\'e dual 2-form
\be
\tw^E = \frac{1}{2}[ R^+ - R^-],
\ee
which represents  the generator of the relative $U(1)$ symmetry associated with the brane-image pair. The observation that $\tw^E$ defines
a non-trivial element of $H^2(Y_E)$ which do not extend to an element of  $H^2(Y_0)$ is in agreement with the fact that this $U(1)$ symmetry is massive \cite{Martucci:2015dxa}.

Over the matter curve ${\cal C}_{II}$, $R^+$ becomes further reducible in that its rational fibre splits into $\mathbb P^1_E \cup \mathbb P^1_+$ with
 \be
 \begin{aligned}
& \mathbb P^1_{E} =   \{w=0\}\cap  \{\chi=0\}  \cap     \{y = 0\} \cap \{s =0 \}\cap \pi^{-1}T\,, \\
 &  \mathbb P^1_{+}  = \{w=0\}\cap  \{\chi=0\}  \cap     \{y+a_1 s = 0\} \cap \{(s \, a_{21} + 2 \, \eta  \, t) \, \lambda_1 + 2 \, \lambda_2 \, a_1 =0 \}\cap \pi^{-1}T\,.
 \end{aligned}
 \ee
Here we have introduced $T$ as some auxiliary base divisor intersecting ${\cal C}_{II}$ in one generic point  to single out one copy of the fibral curve.
Let us furthermore denote the fibre of $R^-$ over ${\cal C}_{II}\cap T$ by
\be
\mathbb P^1_- =\{w=0\}\cap  \{\chi=0\} \cap \{y-a_1 s = 0\}  \cap \{\lambda_1 = 0\}\cap \pi^{-1}T\,  .
\ee
These fibral curves have the following intersection numbers with $R^\pm$,
\be
\begin{aligned}
& \mathbb P^1_E \cdot R^- = 1\,, \qquad & \mathbb P^1_E \cdot R^+ = -1\,, \\
& \mathbb P^1_+ \cdot R^- = 0\,, \qquad & \mathbb P^1_+ \cdot R^+ = 0\,,\\
& \mathbb P^1_- \cdot R^- = -1\,, \qquad & \mathbb P^1_-   \cdot R^+ = 1\,.
\end{aligned}
\ee 
These can be computed by counting the number of transverse intersections in the fibre in an elementary way. Non-effective intersections can be deduced by noting that $R^+ + R^-$ can be deformed away from $W$ because it represents the full fibre over $W$. Hence its intersection number with $\mathbb P^1_\pm$ and $\mathbb P^1_E$ vanishes. Thus e.g. and ${\mathbb P}^1_E \cdot R^+ = -{\mathbb P}^1_E \cdot R^- = -1$. 
From these considerations we read off the $U(1)$ charge of matter from an M2-brane wrapping $\mathbb P^1_E$ as
\be
q = \frac{1}{2} (R^+ - R^-) \cdot \mathbb P^1_E= -1\,.
\ee
For later purposes we observe that also
\be
\frac{1}{2} (R^+ - R^-) \cdot (\mathbb P^1_+ - \mathbb P^1_-)= -1\,.
\ee

According to the above general prescription, a gauge flux with respect to the massive $U(1)$ on the 7-brane wrapping $W$ uplifts on $\hat Y_E$ to 
\be
G_{4}^E = \frac{1}{2} [A^+ - A^-], \qquad A^\pm  = R^\pm \cap {\cal C}_{\rm flux}\,.
\ee
Here ${\cal C}_{\rm flux}$ is a 2-cycle on $W$ whose homology class is Poincar\'e dual to the gauge flux. In general the flux is subject to the non-trivial constraint (\ref{D5tadpole}), which might force us to switch on also gauge flux on the invariant 7-brane.\footnote{If this flux restricts non-trivially to the instanton divisor to be introduced momentarily, its contribution to the Bianchi identity must also be cancelled. We do not describe this explicitly here. }

Consider now the vertical divisor wrapped by the M5-instanton, which is likewise described by a family $D_\epsilon\subset Y_\epsilon$, stably degenerating at $D_0 = \pi^{-1} D_{\rm b}$ on $Y_0$. $D_0$ splits into two rationally fibered components $D_T \subset Y_T$ and $D_E \subset Y_E$ \cite{Clingher:2012rg} in such a way that \be
D_0=D_E\cup_E D_T
\ee
with $E$ the divisor wrapped by the D3-instanton on the Type IIB Calabi-Yau 3-fold $X$.
Notice that only the component $D_E$ is sensitive to the presence of the gauge flux and the 7-branes. The base divisor $D_{\rm b}$ intersects $W$ in an effective curve
 \be
 {\cal C}_{D} = W \cap D_{\rm b}.
 \ee
We are interested in the curve $C_G$ dual to the pullback of the flux to $D$,
\bea
C_G =  \frac{1}{2} (A^+ - A^-) \cap D_E\,.
\eea
 $C_G$ has support in the fibre over the intersection points of the 2-cycles ${\cal C}_{\rm flux}$ and ${\cal C}_{D}$ inside $W$.
   Since the 2-cycle  ${\cal C}_{\rm flux}$ is in general not effective, these intersection points  split into $n^+$ effective intersection points $p_i$ and $n^-$ anti-effective ones $q_j$,
   \be
   {\cal C}_D \cap {\cal C}_{\rm flux}=\sum^{n_+}_{i=1}\{p_i\}-\sum^{n_-}_{j=1}\{q_j\}\ \subset D_{\rm b} .
   \ee 
   The cycle $C_G$ is therefore in general given by the linear combination
 \bea
 C_G = \sum_{i=1}^{n^+} \frac{1}{2} (R^+-R^-)|_{p_i} - \sum_{j=1}^{n^-} \frac{1}{2} (R^+-R^-)|_{q_j}. 
 \eea

 What remains is to understand the fibre of $\frac{1}{2} (R^+-R^-)$ over these points, up to equivalence in perturbative homology. 
  It is instructive to holomorphically transport the fibre along ${\cal C}_D$ from the intersection points ${\cal C}_D \cap {\cal C}_{\rm flux}$ to ${\cal C}_D \cap {\cal C}_{II}$, where the fibre of $R^+$ splits as explained above. 
 Here the fibre of $\frac{1}{2} (R^+-R^-)$ takes the form $\frac{1}{2} (\mathbb P^1_E  + \mathbb P^1_+ -  \mathbb P^1_-)$ and thus
\bea
\Big[C_G  -  \sum_{i=1}^{n_+}   \frac{1}{2} ({\mathbb P}^1_{E} + \mathbb P^1_+ -  \mathbb P^1_-)_i     +  \sum_{j=1}^{n_-}  \frac{1}{2}  ({\mathbb P}^1_{E} + \mathbb P^1_+ -  \mathbb P^1_-)_j  \Big]_{D,{\rm p}}  = 0.
\eea
The subscripts denote that the fibral cycles are equivalent to $\frac{1}{2}(R^+-R^-)|_{p_i}$ or $\frac{1}{2}(R^+-R^-)|_{p_j}$ in perturbative homology.
As noted before,  $\mathbb P^1_+ -  \mathbb P^1_-$ has the same massive $U(1)$ charge as $\mathbb P^1_E$. Away from the stable degeneration limit, we know that the fibre over ${\cal C}_{II}$ is of Kodaira type $I_2$, which leaves room only for a single resolution $\mathbb P^1$. Therefore, $\mathbb P^1_+ -  \mathbb P^1_-$ asymptotes to $\mathbb P^1_E$   on $Y_\epsilon$ with $\epsilon \neq 0$,  the only other possibility being that  it just desappears. In any case, the curve $C_G$ becomes  trivial in the homology of $D_\epsilon$ away from the stable degeneration point and thus meets the crucial prerequisite that it can be canceled by the instanton flux.
More precisely, one can  adapt the discussion in \cite{Martucci:2015dxa} and argue for the presence of a non-perturbative 3-chain $\Gamma^{\rm flux}_{\rm np}\subset D$ with 
\be
\partial \Gamma^{\rm flux}_{\rm np} = C_G,
\ee
Then the contribution of $C_G$ to the Bianchi identity can be canceled via a $T_3$ flux such that
\be
\d T_3 = \delta_4(C_G).
\ee
If such $T_3$ flux satisfies the supersymmetry conditions,  the M5-brane instanton produces an F-term $\sim e^{-S_{\rm M5}}$  not containing any charged matter insertion.

\section{$U(1)$ enhancement along instantons}
\label{sec:gaugenh}

 We now come back to the interesting phenomenon pointed out in section \ref{sec_pertvsnonpert} that a vertical divisor $D$ can contain extra sections which do not arise by pullback from the ambient space $\hat Y$. 
Generally in such a situation, the M5-brane on $D$ probes a `locally massless $U(1)$' gauge group which is broken globally on $\hat Y$. Physically, the Higgs field responsible for the mass of the $U(1)$ potential 
has vanishing restriction to $D$. This could be either a Higgs field localised on a matter curve or another axionic field participating in a geometric St\"uckelberg mechanism for the $U(1)$. See \cite{Martucci:2015dxa} for a distinction of these two. In models with a well-defined weak-coupling limit, the latter situation corresponds to vanishing restriction of the odd 2-cocycle associated with the $C_2$ axion on the D3-brane instanton divisor $E\subset X$. 

Apart from being interesting by itself, the appearance of extra sections on M5-instantons implies a subtle selection rule affecting the instanton contributions to the path integral which must be taken into account. The reason is that in this case the pullback of homologically trivial fibral curves can be homologically non-trivial on $D$. This may obstruct an otherwise expected contributions to the effective action. As a warm-up we first describe the appearance of a `local $U(1)$' along an M5-instanton in a simple example, and then apply the same observations to a fibration with a multi-section.

\subsection{A `local $U(1)$' along a vertical divisor}

Consider an elliptic 4-fold $Y$ given by a Tate model of the form $P=0$ with
\bea \label{plainTateY}
P = x^3  - y^2  - a_1 \, x \, y \, z  - a_3 \, y \, z^3  +  a_2 \, x^2 \, z^2 + a_4 \, x \, z^4 + a_6 \, z^6,
\eea
where $a_i$ are sections of $\bar K^i$. Assume that $a_6 = \rho \, \tau$, with $\rho$ and $\tau$ sections of appropriate line bundles. 
Imagine now an M5-brane instanton along the divisor $D = \pi^{-1} D_{\rm b}$, with  $D_{\rm b} = \{ \tau = 0\}$ being a single connected effective divisor. The vertical divisor $D$ is by itself an elliptically fibered 3-fold described by 
\bea
P_D = x^3  - y^2  - \tilde a_1 \, x \, y \, z  - \tilde a_3 \, y \, z^3  +  \tilde a_2 \, x^2 \, z^2  + \tilde a_4 \, x \, z^4
\eea
with $\tilde a_i:= a_i|_{D_b}$ sections of $\tilde{\bar K}^i = {\bar K^i}|_{D_b}$. 
Since $\tilde a_6 \equiv 0$, this is just a $U(1)$ restricted Tate model \cite{Grimm:2010ez}. Indeed the fibration $D$ has another rational section ${\rm Sec}_D:  [ x   : y  : z] = [0 : 0 : 1]$ independent of the holomorphic zero-section $Z_D: [ x   : y  : z] = [1 : 1 : 0]$. The latter is just the pullback of the holomorphic zero-section $Z$ on $Y$. The divisor $D$ is singular as a 3-fold with an $I_2$-singularity located at 
 $x=y=\tilde a_3 = \tilde a_4 = 0$. This can be seen from the vanishing orders of $(\tilde a_1, \tilde a_2, \tilde a_3, \tilde a_4, \tilde a_6; \Delta) = (0,0,1,1,\infty;2)$, where the entry $\infty$ denotes that the corresponding section vanishes identically. Note that these points are not singularities of the ambient 4-fold $Y$ because at $a_3=a_4=\tau=0$ the Tate sections vanish as $( a_1, a_2, a_3, a_4, a_6; \Delta) = (0,0,1,1,1;1)$.

Since the rational section ${\rm Sec}_D$ passes through the singularity in the fibre it becomes singular itself. In order to avoid having to discuss the (co)homology of $D$ as a singular space, we can resolve it by a small resolution or by a standard blow-up $(x,y) \rightarrow (x \, s,y \, s)$. The latter option leads to an auxiliary space $\hat D$ given by the proper transform of $D$ with respect to the blow-up,
 \bea
P_{\hat D} = x^3 \, s^2 - y^2 \, s - \tilde a_1 \, x \, y \, z \, s - \tilde a_3 \, y \, z^3  +  \tilde a_2 \, x^2 \, z^2 \, s + \tilde a_4 \, x \, z^4.
\eea
The fibre over the point set $\{\tilde a_3 = 0\} \cap \{\tilde a_4=0\}$ in $\hat D$ splits into two rational curves intersecting like a Kodaira $I_2$-fibre. We will denote by  $\mathbb P^1_S$ the resolution $\mathbb P^1$ which replaces the singular fibre point $x=y=0$.

We should stress that the transition to $\hat D$ does not mean that we perform an actual blow-up on $Y$. Rather we define the resolution $\hat D$ as an auxiliary space which asymptotes to $D$ in the F-theory limit of vanishing fibre volume. We will see momentarily how to relate the topological data of $\hat D$ to observables of the physical F-theory model.




The resolved rational section is identified with the divisor $S_{\hat D}= \{s=0\}$ on ${\hat D}$.  
As in the $U(1)$ restricted Tate model this divisor wraps the resolution $\mathbb P^1_S$ over $\{\tilde a_3 = 0\} \cap \{\tilde a_4=0\}$ in $\hat D$.
This makes the appearance of a `local massless $U(1)$' along $\hat D$ (and thus also on $D$) evident, generated by the combination $S_{\hat D}-Z_{\hat D}$. From a type IIB perspective the reason for this local enhancement is easy to understand.
Recall first that the Sen limit of a $U(1)$ restricted Tate model, i.e. of a 4-fold $Y$ of the form (\ref{plainTateY}) with $a_6 \equiv 0$ globally, corresponds to a brane-image-brane pair along two homologous divisors intersecting on a curve on the Calabi-Yau double cover $\X$.\footnote{In fact, the two branes intersect on two curves, one of which lies on top of the O7-plane. We neglect this intersection curve since no massless matter is localised here.} This curve is the uplift of the curve $\calc_{34} = \{a_3 =0\} \cap \{ a_4=0\}$ to $\X$ \cite{Grimm:2010ez}. The $U(1)$ gauge group is the relative $U(1)$ between both D7-branes. Deforming this to the generic Tate model by switching on $a_6$ breaks this $U(1)$ completely by moving along the Higgs branch for the charged matter fields along ${\cal C}_{34}$. This recombines the two D7-branes into an invariant D7-brane of Whitney-umbrella type \cite{Collinucci:2008pf}.
Due to the specialisation of $a_6 = \rho \, \tau$ this divisor takes a special form \cite{Braun:2011zm}. The instanton divisor at $\tau=0$ intersects the Whitney brane in two separate, intersecting curves exchanged by the orientifold involution. I.e.\ the restriction of the invariant 7-brane to the instanton splits up into a brane-image-brane pair. This is the reason why the elliptic fibration $D$ itself takes the form of a $U(1)$ restricted Tate model. The `local $U(1)$' probed by the instanton is the associated relative $U(1)$ which is unbroken only on the intersection of the 7-brane with the instanton divisor, and the $I_2$-singularities on $D$ sit over the uplift of the intersection points of the two intersecting curves within $\{\tau=0\}$. Physically, the Higgs field which is responsible for the breaking of $U(1) \rightarrow \emptyset$ in the ambient space restricts to zero locally on the divisor $\{\tau = 0\}$.

In the Tate model (\ref{plainTateY}) with $a_6 = \rho \,  \tau$, one can consider the 4-form flux \cite{Braun:2011zm}
\bea
G_4 = \Sigma_\rho -  (Z + \pi^{-1} \bar K) \cdot \pi^{-1}{\cal P}     \in {\rm CH}^{2}(Y).
\eea
Here
\bea
\Sigma_\rho = \{x=0\} \cap \{y=0\} \cap \{\rho=0\} \subset Y_5
\eea
describes a 4-cycle in the ambient space $Y_5$ of $Y$ which automatically lies on $Y$ due to the factorisation  $a_6 = \rho \, \tau$. Furthermore we have defined ${\cal P} = \{\rho = 0\}$.
In fact, in the presence of this flux the restriction $a_6 = \rho \,  \tau$ follows dynamically by solving the F-terms resulting from the induced superpotential \cite{Braun:2011zm}.
On the singular divisor $D \subset Y$ this flux $G_4$ restricts to
\bea
G_4|_D = ({\rm Sec}_D - Z_D - \pi^{-1}\tilde{\bar K}) \cdot \pi^{-1} \tilde {\cal P}
\eea
with $\tilde {\cal P} = {\cal P} |_{D_{\rm b}}$.
As discussed above, on the auxiliary space $\hat D$, the independent singular section ${\rm Sec}_D$ is resolved into $S_{\hat D}$  and we are to consider
\bea
\hat G_4 := (S_{\hat D} - \tilde Z_{\hat D} - \pi^{-1}\tilde{\bar K}) \cdot \pi^{-1}\tilde {\cal P} \in {\rm CH}^2(\hat D). 
\eea
In particular we see that $\hat G_4$  is cohomologically non-trivial on $\hat D$ (and, in fact, the same is true for $G_4|_D$ on the singular space $D$). Therefore it is not possible to cancel its contribution to the Bianchi identity in terms of $T_3$ alone. 
To compute the homology class of its dual curve $C_G$, recall that the fibre over $\{\tilde a_3=0\} \cap \{\tilde a_4=0\}$ splits into two intersecting $\mathbb P^1$s.
By comparison of intersection numbers, using in particular that
\bea
\int_{\mathbb P^1_S} S_{\hat D} = -1, \qquad \int_{\mathbb P^1_S} Z_{\hat D} = 0,
\eea
one can show that 
\bea
[C_G] = n \,  [\mathbb P^1_S], \qquad n= 2 \int_B {\cal P} \wedge D_{\rm b} \wedge {\bar K} . 
\eea
To satisfy the Bianchi identity it is therefore necessary to insert M2-branes wrapping suitable copies of $\mathbb P^1_S$ such as to cancel this contribution.
From the perspective of the `local $U(1)$' on $\hat D$, these states have charge $\pm 1$. However, on $Y$ no such notion of a $U(1)$ exists. 
Rather, these states should be interpeted as uncharged brane moduli as opposed to localised charged matter. 
Indeed, if we set $a_6 \equiv 0$ globally on $Y$, the fibre over the curve $\{a_3 =0\} \cap \{a_4=0\}$ is singular and charged matter of charge $\pm 1$ resides here. As recalled before the deformation to $a_6 \neq 0$ corresponds to a D-flat Higgsing. E.g. if there is precisely one vectorlike pair $(\Phi_1, \tilde \Phi_{-1})$ of localised Higgs fields, one linear combination of ${\cal N}=1$ chiral mulitplets is absorbed by the super-Higgs mechanism and one chiral superfield remains as a brane modulus. The profile of this brane modulus is delocalised along the 7-brane. But on $D$, where the Higgs VEV vanishes, we can locally describe it in terms of  an M2-brane wrapping the resolution curve $\mathbb P^1_S$ after moving to the resolved space $\hat D$. This reflects the origin of this field as a combination of charged modes before Higgsing.

This interpretation is in agreement with the behaviour of the brane-image configuration in the Sen limit. 
A Whitney brane on a divisor  $W$ intersects a \emph{generic} invariant divisor $E$ in a single curve $C_{WE} = W \cap E$, which is invariant under the orientifold involution. Since the gauge flux $F_W$ is orientifold odd, $\int_{C_{WE}} F_W = 0$. Zero modes localised on $C_{WE}$ are, if any, non-chiral. They can be absorbed by couplings of the form $\lambda \Phi \tilde \lambda$ with $\Phi$ a deformation modulus of $W$. 
For the specific divisor associated with our instanton however, $C_{WE}  = C_1 \cup C_2$ with $\sigma_*(C_1) = C_2$. The statement that $[G_4|_D] \neq 0 \in H^4(D)$ simply means that $\int_{C_1} F_W = - \int_{C_2} F_W \neq 0$.
Since no global $U(1)$ exists this does still not imply an actual chiral excess of brane-instanton modes. Put differently, a mode $\lambda$ stretching from $E$ to $W$ on $C_1$ is paired with a mode $\tilde \lambda$ stretching form $W$ to $E$ on $C_2$, with which it is identified under the orientifold action. The zero modes can be absorbed by the same type of coupling to a brane modulus as before.

\subsection{Consequences for matter couplings in a multi-section fibration} \label{sec_Z2instanton}

Since  the plain Tate model (\ref{plainTateY}) exhibits no localised matter, this setup is not appropriate to illustrate the consequences of a `local $U(1)$' for the contributions of the M5-instanton to the charged matter couplings. 
A generalisation of this setup is to allow for an $n$-section on the ambient 4-fold $Y$ to split on $D$ into $n$ independent sections. This happens whenever $D$ misses the branching locus on the base around which the $n$ intersection points of the multi-section with a local fibre are exchanged. 
A simple example is the bisection model discussed at length in \cite{Braun:2014oya,Morrison:2014era,Anderson:2014yva,Klevers:2014bqa,Garcia-Etxebarria:2014qua,Mayrhofer:2014haa,Mayrhofer:2014laa} (see \cite{Cvetic:2015moa} for the discussion of a  trisection model) given by the hypersurface constraint $P=0$ with
\be \label{bisectionfibration}
P:=  \, w^2   + b_{0}   \, u^2 \,  w  + b_1 \, u \, v \, w  + b_2  \, v^2 \, w + c_0  \, u^4   + c_1   \, u^3 \, v  + c_2  \, u^2 \, v^2   + c_3 \, u \, v^3 + c_4 \, v^4\,.
\ee
It has a bisection $U:  \{u=0\} \cap \{ w + \frac{1}{2} v^2 (b_2 \pm \sqrt{b_2^2 - 4 c_4}) = 0 \}$. 
Suppose again that $c_4 = \rho \, \tau$ and consider an M5-instanton along the vertical divisor $D= \pi^{-1} D_{\rm b}$ with $D_{\rm b} = \{\tau= 0\}$. The 3-fold $D$ is an elliptic fibration $P_D=0$ with
\be
P_D:=  \, w^2   + b_{0}   \, u^2 \,  w  + \tilde b_1 \, u \, v \, w  + \tilde b_2  \, v^2 \, w + \tilde c_0  \, u^4   + \tilde c_1   \, u^3 \, v  + \tilde c_2  \, u^2 \, v^2   + \tilde c_3 \, u \, v^3
\ee
where $\tilde b_i = b_i|_{D_{\rm b}}$ etc.
The restriction of the bisection $U|_D$ splits into two local sections on $D$ given by
\be
U_D: [u  : v :  w] = [0 : v : 0]\quad,     \qquad S_D: [u : v : w] = [0 : v : -b_2 v^2] \,.
\ee
By the same reasoning as in our first example, this is because setting $c_4 \equiv 0$ unhiggses the $\mathbb Z_2$ discrete symmetry present in the F-theory compactification to a massless $U(1)$ gauge symmetry \cite{Morrison:2014era}. 
The divisor $D$ acquires a conifold singularity in the fibre over $\{\tilde b_2=0\} \cap \{\tilde c_3=0\}$ while the divisor $D_{\rm b}$ (and also the ambient 4-fold) is smooth. 
As before we consider as an auxiliary space the resolution of $D$ by blowing up $(u,w)\rightarrow (u \, s, w\, s)$ and passing to the proper transform $\hat D=\{P_{\hat D}=0\}$ with
\be 
P_{\hat D} =s  \, w^2   + \tilde b_{0}  \, s^2 \, u^2 \,  w  + \tilde b_1 \, s \, u \, v \, w  + \tilde b_2  \, v^2 \, w + \tilde c_0 \, s^3 \, u^4   + \tilde c_1 \, s^2  \, u^3 \, v  + \tilde c_2 \, s \, u^2 \, v^2   + \tilde c_3 \, u \, v^3\,.
\ee
This is the $U(1)$ model of \cite{Morrison:2012ei}, which generalises the previous $U(1)$ restricted Tate model. 
The two independent sections can be identified with
\be
U_{\hat D} = \{u=0\}, \qquad S_{\hat D} = \{s=0\}\,.
\ee

Let us now turn to the effect of this instanton on matter couplings.
On the 4-fold $Y$ the torus fibre splits, over a certain curve $\calc_{II}$,  into two \emph{homologous} rational curves $A_{II}$ and $B_{II}$.\footnote{In \cite{Mayrhofer:2014laa,Martucci:2015dxa}, to which we refer for details, these objects had been called $\tilde C_{II}$, $\tilde A_{II}$ and $\tilde B_{II}$ to distinguish them from analogous objects in the $U(1)$ model with $c_4 =0$. We are dropping the tilde here for notational convenience.} In fact, for a fibration over a generic base space, $[A_{II} - B_{II}]_{\rm p} = 0$ \cite{Martucci:2015dxa}.
This is because both $\mathbb P^1s$ intersect the bisection $U$ in one point each, but these two points are exchanged globally by a monodromy around a set of points given by the intersection of $\calc_{II}$ with the branching-locus $\{b_2^2 - 4 c_4=0\} $ of the bisection. In absence of this point set, we merely have $[A_{II} - B_{II}]_{\rm np} = 0$. This can happen for special base spaces with suitable intersection numbers as specified in \cite{Martucci:2015dxa}.
On $\hat D$, as a result of the splitting of the bisection into $U_{\hat D}$ and $S_{\hat D}$, $[A_{II} - B_{II}]_{D,{\rm np}} \neq 0$. Indeed, under the `local $U(1)$' generator $S_{\hat D} - U_{\hat D}$ M2-branes wrapping both curves have opposite charges and are thus distinguishable. 

Of special interest is now the situation in which the set of monodromy points is indeed absent on $\Y$ and thus $[A_{II} - B_{II}]_{\rm np} = 0$, but $[A_{II} - B_{II}]_{\rm p} \neq 0$. 
This means that an operator $\tilde \Phi_{A_{II}} \Phi_{B_{II}}$ (plus conjugate) cannot exist perturbatively on $Y$ \cite{Martucci:2015dxa}.
The only perturbative operators are the trivially possible terms $\tilde \Phi_{A_{II}} \Phi_{A_{II}}$ and $\tilde \Phi_{B_{II}} \Phi_{B_{II}}$.\footnote{The latter are due to an ingoing and outgoing M2 on $A_{II}$ (or $B_{II}$), while an operator $\tilde \Phi_{A_{II}} \Phi_{B_{II}}$ alone would require an ingoing M2 on $A_{II}$ and an outgoing one on $B_{II}$. This is not possible if $[A_{II} - B_{II}]_{\rm p} \neq 0$.}
From the perspective of the 4-dimensional F-theory effective action, $\Phi_{B_{II}}$ and $\tilde \Phi_{ A_{II}}$ both correspond to the same ${\cal N}=1$ chiral superfield $\Phi_{II}$. 
To determine the number of perturbatively massless states one must count the zero modes on $ C_{II}$. On top of this perturbative computation, an M5-instanton along a vertical divisor along which $c_4|_D \neq 0$ can contribute an additional, non-perturbative mass term
provided it supports supersymmetric 3-form flux such that 
\be \label{BianchiAB1}
\d T_3 = - \delta_{4}(A_{II} - B_{II})\,.
\ee
Here it is crucial that indeed $[A_{II} - B_{II}]=0 \in H_2(D)$ for generic $D$ such that this solution can exist. 
More precisely, if a fluxed instanton satisfying the Bianchi identity (\ref{BianchiAB1}) contributes to the superpotential, it generates there a non-perturbative mass term $\Phi_{II} \Phi_{II} \,   e^{-S_{M5}}$ which in particular can be interpreted as a Majorana mass for the fermionic perturbative zero modes.

By contrast our special instanton divisor with $D_{\rm b} = \{\tau = 0\}$  cannot fulfill the condition (\ref{BianchiAB1}). 
This superselection rule can only be bypassed in a background with suitable gauge flux on $Y$  whose restriction to $\hat D$ is likewise cohomologically non-trivial.
In \cite{Mayrhofer:2014haa} a gauge flux for the bisection fibration (\ref{bisectionfibration}) was proposed of the form
\be \label{bisection-flux}
G_4 = \sigma_0 - \sigma_1 \in {\rm CH}^2( Y)
\ee 
for
\be
\sigma_0 = \{u=0\} \cap \{w=0\} \cap \{\rho = 0\}, \qquad \sigma_1 = \{u=0\} \cap \{w=-b_2 v^2\} \cap \{\rho = 0\}.
\ee

To compute  $[G_4|_{\hat D}] \in H^4(\hat D)$ first note that
$[G_4|_{\hat D}] = [(U_{\hat D} - S_{\hat D})] \wedge [ \pi^{-1} \tilde {\cal P}]$ with $\tilde {\cal P} ={\cal P}|_{D_{\rm b}}$ and ${\cal P} = \{\rho=0\}$. We can then make an ansatz for the class of the Poincar\'e dual 2-cycle 
$C_G$.
Recall furthermore that $\hat D$ is a Bl$^1 \mathbb P_{112}[4]$ fibration of the form studied in \cite{Morrison:2012ei}. We can therefore 
use our knowledge of the intersection numbers and the structure of fibral curves of this model, as briefly summarised for our purposes e.g.\ in \cite{Mayrhofer:2014haa}.
Apart from the resolved $I_2$ fibre with components $ A_{II}$ and $ B_{II}$, the fibre of  $\hat D$ hence degenerates at the points $\{\tilde b_2 = 0\} \cap \{\tilde c_3 = 0\}$ into two intersecting $\mathbb P^1s$ $ A_I$ and $ B_{I}$. 
The intersection structure of fibral curves and sections on $\hat D$ is
\be
\begin{aligned}
& U_{\hat D} \cdot A_I = 1, \qquad S_{\hat D} \cdot A_I = -1,  \qquad U_{\hat D} \cdot B_I = 0, \qquad S_{\hat D} \cdot B_I = +2, \\
& U_{\hat D} \cdot A_{II} = 1, \qquad S_{\hat D} \cdot A_{II} = 0,  \qquad U_{\hat D} \cdot B_{II} = 0, \qquad S_{\hat D} \cdot B_{II} =1.
\end{aligned}
\ee
We can now make the ansatz $[C_G] = a [A_I] + b [B_I] + [\gamma_{\rm b}]$ with $\gamma_{\rm b}$ any curve on $D_{\rm b}$. 
With the help of the intersection numbers
\be
\int_{\hat D} S_{\hat D} \wedge U_{\hat D} \wedge \pi^{-1} \tilde {\cal P} = \int_{D_{\rm b}} \tilde b_2 \wedge \tilde {\cal P}, \qquad \int_{\hat D} U_{ \hat D} \wedge U_{\hat D} \wedge \pi^{-1}\tilde {\cal P} = \int_{D_{\rm b}} (- \tilde{\bar K}) \wedge \tilde {\cal P} = \int_{\hat D} S_{\hat D} \wedge S_{\hat D} \wedge \pi^{-1} \tilde {\cal P}
\ee
with $\tilde{\bar K} = {\bar K}|_{D_{\rm b}}$ one can compare this ansatz with the intersection of $G_4|_{\hat D}$ with $U_{\hat D}$, $S_{\hat D}$ and any pullback divisor from $D_{\rm b}$. This yields 
\be
[C_G] = n \, [A_I] = n \,  [A_{II} - B_{II}],  \qquad n = \int_{D_{\rm b}} (\tilde b_2 + \tilde{\bar K}) \wedge \tilde {\cal P}.
\ee
In particular there is no cohomological obstruction to solving the Bianchi identity
\be\label{t3n}
\d T_3 = \delta_4(C_G) - n \, \delta_4(A_{II}) +n\, \delta_4(B_{II}).
\ee
To determine whether $T_3 \neq 0$ is required one must study $C_{G}$ at the level of rational equivalence classes rather than merely compute its cohomology class on $\hat D$.
A supersymmetric $T_3$ satisfying (\ref{t3n}) can then  give rise to an F-term of the schematic form
\be
 (\Phi_{II} \Phi_{II})^n  e^{-S_{M5}}.
\ee
 This can arise at the level of the superpotential if the appropriate conditions on the fermionic zero modes on $D$ are satisfied.




\section{Conclusions}\label{sec_Conclusions}

We have analysed the selection rules governing the contribution of M5-brane instantons to the effective action of general F-theory compactifications.
Our starting point has been the Bianchi identity (\ref{T3BI}), which relates the classical M5-instanton flux $T_3$ to the pullback of the 4-form background gauge flux $G_4$.
If this equation has a solution requiring in addition the insertion of time-like M2-branes on suitable curves in the torus fibre, then the fluxed M5-instanton gives rise to an operator in the effective action dressed by the associated charged matter fields \cite{Donagi:2010pd,Marsano:2011nn,Kerstan:2012cy}. This is the F/M-theory analogue of by now well-known mechanisms for D-brane or heterotic worldsheet instantons in weakly coupled string theories \cite{Blumenhagen:2006xt,Ibanez:2006da,Florea:2006si,Haack:2006cy}. Such charged matter dependence of the instanton generated coupling can be a curse or a blessing: The latter because the resulting non-perturbatively suppressed operators can be phenomenologically interesting by themselves (see e.g. \cite{Blumenhagen:2009qh} and references therein), the first because this can be a challenge to the stabilisation of all K\"ahler moduli \cite{Blumenhagen:2007sm}. It is therefore of considerable interest to advance our understanding of  the M5-instanton selection rules in F-theory.

Progress in this direction hinges upon the central question in what geometric sense the Bianchi identity  (\ref{T3BI}) must be satisfied.
We have argued that the specific  instanton contribution to the partition function is encoded in an analysis of the Bianchi identity modulo perturbative homological relations as introduced in our companion paper \cite{Martucci:2015dxa}. Perturbative homological relations are defined as relations in the fibre homology up to deformations which holomorphically transport fibral curves along 3-chains with one leg in the base only. For instance, rationally equivalent fibral cycles are also equivalent in this sense. This makes contact with the description of background gauge fluxes in terms of rational equivalence classes of 4-cycles \cite{Bies:2014sra}.

Our approach enables us in particular to investigate situations in which the pullback of the  flux $G_4$ to the instanton divisor is trivial in the homology of the instanton divisor $D$, but non-trivial as a rational equivalence class on $D$. In this case gauge invariance with respect to massless $U(1)$ gauge symmetries poses no constraints on the instanton contributions, yet we have seen how the presence of the gauge flux must - and can - be compensated by insertion of suitable time-like M2-branes and/or by a supersymmetric 3-form field strength $T_3$ on the instanton. Such fluxed M5-instantons can be viewed as M5-M2-bound states which, unlike the M2-instantons themselves, can contribute to the superpotential or other F-terms. From an effective field theory point of view, the selection rules we have found in this case can be interpreted as due to a geometrically massive $U(1)$ symmetry in the sense of \cite{Grimm:2010ez,Grimm:2011tb}. We have been able to find an intrinsic F/M-theoretic description of such effects which agree, in models with a weak coupling limit, with results from fluxed D3-brane instantons \cite{Grimm:2011dj,Bianchi:2011qh}.

In this context we have also revisited the challenge of constructing the gauge flux associated with massive $U(1)$ symmetries in F-theory. We have extended the analysis of \cite{Braun:2014nva}  and investigated such flux 
in the stable degeneration limit \cite{Donagi:2012ts,Clingher:2012rg} of a prototypical massive $U(1)$ model. This way we have been able to confirm expectations from the IIB limit concerning the interplay of massive $U(1)$ fluxes and instanton flux. A mathematical challenge for the future will be to extend this description of massive $U(1)$ flux away from the stable degeneration limit to non-weakly coupled F-theory compactifications. 
A crucial role in this context will be played by the logarithmic cohomology of the stably degenerate 4-fold arising in the Sen limit \cite{Donagi:2012ts,Clingher:2012rg}.

From a formal perspective it would be desirable to gain a deeper understanding of the relation between the type of homological relations considered in our work and standard notions of equivalence between cycles such as algebraic and rational equivalence. A related  challenge is to explicitly construct the 3-chains  and the dual supersymmetric non-closed 3-form flux on the instanton appearing in the Bianchi identity. Furthermore, our conclusions regard the selection rules governing the qualitative structure of the couplings produced in the effective actions. Much more effort would be required to actually identify the precise form of such terms, by computing the instanton partition function. This important missing step may be achieved by combining the F/M-theory description adopted in the present paper with the purely non-perturbative IIB approach proposed in \cite{Bianchi:2011qh,Bianchi:2012kt,Martucci:2014ema}, in which the instanton effective action is described by a four-dimensional duality twisted  $N=4$ SYM  theory.

An interesting phenomenological off-spring of our analysis is the observation that fluxed M5-instantons do in general produce non-perturbative corrections also to perturbatively allowed Yukawa couplings beyond the ones considered so far in the literature  \cite{Marchesano:2009rz,Aparicio:2011jx,Font:2012wq,Font:2013ida,Marchesano:2015dfa}. The relevance of this point is that such non-perturbative corrections are expected to modify the rank of the Yukawa couplings e.g. in realistic GUT models and might thus account for the observed hierarchies in the flavour sector. It will be exciting to investigate the consequences of the fluxed instanton contributions proposed in this work in this regard.

\bigskip

\centerline{{\bf Acknowledgements}}

\noindent We thank  Martin Bies, Andreas Braun, Andres Collinucci, Hans Jockers, Ling Lin, Christoph Mayrhofer, Eran Palti, Oskar Till and Roberto Valandro for important discussions and comments. 
This work was partially funded by DFG under Transregio 33 `The Dark Universe' and by
the Padua University Project CPDA144437.

\bigskip

\appendix

\section{Type IIB backgrounds at weak coupling} \label{chargedD3inst}

Consider a general weakly coupled IIB  orientifold on a Calabi-Yau 3-fold   $\X$ with D7-branes and $O7$-planes associated with a holomorphic orientifold involution $\sigma:\X\rightarrow \X$. The D7-branes can appear either in pairs $({\rm D7}_a,{\rm D7}'_a)$,  wrapping internal effective divisors (i.e.\ holomorphic 4-dimensional submanifolds) $(D_a,D'_a)$  with $D'_a=\sigma_*D_a\neq D_a$, or as single D7$_\alpha$, wrapping orientifold invariant effective divisors $\hat D_\alpha=\sigma_*\hat D_\alpha$. In this section, in order to simplify the presentation, we also assume that there are no stacks of multiple D7-branes, which could be straightforwardly included in the discussion. The field-strengths on the D7-branes are denoted by $F^a$, $F^{a\prime}$ and $\hat F^\alpha$ respectively, which must satisfy the orientifold projections $F^{a\prime}=-\sigma^*F^a$ and 
$\hat F^\alpha=-\sigma^*\hat F^\alpha$. In particular, $\hat F^\alpha$ must be odd. We allow for possible non-trivial supersymmetric  expectation values for such field-strengths, i.e.\ such that $F^{0,2}=0$ and $F\wedge J=0$. Such supersymmetric fluxes define corresponding primitive  cohomology classes in $H^{1,1}(D_a)$
and $H^{1,1}(\hat D_\alpha)$. We then denote by $\Sigma_a$, $\Sigma'_a$ and $\hat\Sigma_\alpha$ the 2-cycles in $D_a$, $D'_a$ and $\hat D_\alpha$ that are Poincar\'e dual to $\frac{1}{2\pi}F^a$, $\frac{1}{2\pi}F^a{}'$ and $\frac{1}{2\pi}\hat F^\alpha$ respectively. Notice that we are implicitly ignoring a possible half-integer shift of the D7-brane fluxes due to the Freed-Witten quantisation condition \cite{Minasian:1997mm,Freed:1999vc}, which can be taken into account without substantially changing  our discussion. The supersymmetry condition $F^{0,2}=0$ for the D7-brane fluxes can be then restated in homological terms by choosing the dual 2-cycle $\Sigma$ as a linear combination of holomorphic curves, $\Sigma=\calc-\tilde\calc$,  while the primitivity condition is solved by requiring  $\int_{\calc}J=\int_{\tilde\calc}J$. Since  $\int_{\calc}J\geq 0$ and $\int_{\tilde\calc}J\geq 0$ (in absence of singularities), 
we conclude that  both $\calc$ and $\tilde\calc$ must be non-vanishing. 

Once the backreaction of these world-volume fluxes  is taken into account, they generically source a $G_3$ field strength. On top of this, one can also have some purely bulk  $G_3$-flux. However,  our main considerations do not depend on such bulk flux and in this section we  set it to zero for  technical simplicity. Furthermore, we assume that the Neveu-Schwarz field $B_2$ does not contain any half-integer closed contribution that is even under the orientifold involution, a possibility which is a priori possible in perturbative IIB theory but whose F-theory uplift is more involved. 

For the present paper, an important observation is that the four-dimensional $U(1)_a$ gauge symmetries can be massive by a St\"uckelberg mechanism \cite{Jockers:2004yj,Plauschinn:2008yd}.
This follows concretely from the transformations of the Ramond-Ramond (R-R) gauge potentials 
\begin{subequations}\label{RRgaugings}
\begin{align}
C_2 &\rightarrow  C_2+\frac{1}{2\pi}\sum_a\lambda^a\, \Big[\delta_{2}(D_a)- \delta_{2}(D'_a)\Big], \label{RRgaugingC2}\\
C_4&\rightarrow  C_4- \frac{1}{4\pi^2}\sum_a\lambda^a\Big[F^a\wedge \delta_{2}(D_a)- F^{a\prime}\wedge\delta_{2}(D'_a)\Big]\label{RRgaugingC4}
\end{align}
\end{subequations}
under a four-dimensional $U(1)_a$ transformation $A^a \rightarrow A^a + \d \lambda^a$. A quick derivation can be found e.g.\ in Appendix B of \cite{Martucci:2015dxa},  which also explains the notation for the R-R potentials adopted here. The four-dimensional effective theory only sees the cohomological content of (\ref{RRgaugings}), which is given by
\begin{subequations}\label{RRg}
\begin{align}
[C_2] &\rightarrow  [C_2]+\frac{1}{2\pi}\sum_a\lambda^a\, [D_a-D'_a], \label{RRC2}\\
[C_4]&\rightarrow  [C_4]- \frac{1}{2\pi}\sum_a\lambda^a[\Sigma_a-\Sigma_a'],\label{RRC4}
\end{align}
\end{subequations}
where $[D_a-D'_a]$ and $[\Sigma_a-\Sigma_a']$ denote the 2- and 4-cocycles which are Poincar\'e dual to $D_a-D'_a$ and $\Sigma_a-\Sigma_a'$ in $\X$.\footnote{In this paper we often use the same notation $[S]$ for a cycle with representative $S$ and for its Poincar\'e dual cocycle, leaving Poincar\'e duality understood.}

The gaugings (\ref{RRC2}) and (\ref{RRC4}) induce the so-called {\em geometric} and {\em flux-induced} St\"uckelberg masses, respectively, for the $U(1)$ gauge  
transformations. Notice that, since $D'_a=\sigma_* D_a$ and $\Sigma'_a=-\sigma_*\Sigma_a$, $D_a-D'_a$ identifies an orientifold {\em odd} class in $H_4(\X,\mathbb{Z})$ while $\Sigma_a-\Sigma_a'$ identifies an orientifold {\em even} class in $H_2(\X,\mathbb{Z})$. The geometric and flux-induced St\"uckelberg
mechanisms are at work if these particular classes are non-vanishing.  

\subsection{Charged D3-brane instantons at weak coupling}

A supersymmetric instantonic D3-brane must wrap an effective divisor $E\subset \X$ and can support a self-dual world-volume flux $\calf=\frac{1}{2\pi}F_{E}-\iota^* B_2$, $\calf=*_E\calf$, as discussed for instance in \cite{Grimm:2011dj,Bianchi:2011qh,Bianchi:2012kt}. Notice that in general one must sum over all consistent flux configurations in order to evaluate the complete D3-instanton contribution to the four-dimensional effective theory. We focus for simplicity on so-called $O(1)$ instantons \cite{Argurio:2007qk,Argurio:2007vqa,Bianchi:2007wy,Ibanez:2007rs}. This means that $E$ is a single connected orientifold invariant divisor, albeit not pointwise, i.e.\ $E=\sigma_*E$. In this case  the instanton flux must be orientifold odd, $F_E=-\sigma^*F_E$.\footnote{In order to contribute to the superpotential, as opposed to higher order F-terms, the divisor $E$ must be rigid. For more details see e.g. the review \cite{Blumenhagen:2009qh} and references therein. Our considerations can furthermore be adapted to more general D3-brane instantons. In particular, a so-called $U(1)$ instanton 
wraps the pair of divisors $E$ and $E'=\sigma_*E\neq E$. The contribution of such instantons to  the effective theory is more subtle.} The F-term contribution of a single fluxed $O(1)$ instanton to  the effective theory is  proportional to $e^{-S_{\rm D3}}$ with 
\be \label{SD3conventions}
S_{\rm D3}=\frac{\pi}{g_s}{\rm vol}_{\rm DBI}(E) - \pi\ii\int_E C\wedge e^{{\cal F}},
\ee
where the DBI-volume is measured in string frame. Notice that we have included a prefactor of $\frac12$ due to the orientifold projection. 

Let us denote by $\Sigma_E\subset E$ the two-cycle Poincar\'e dual to $\frac{1}{2\pi}F_E$ in $E$, as well as its push-forward to the bulk $\X$.
Notice that the Freed-Witten quantisation condition \cite{Minasian:1997mm,Freed:1999vc} requires, for vanishing discrete background $B_2$-field, $\frac{1}{2 \pi} F_E + \frac{1}{2} c_1(K_E)$ to be an integer class. For a holomorphic divisor $E$ on a Calabi-Yau $X$, $K_E = E$. Since $F_E$ is orientifold odd but $E$ is, by assumption of an (orientifold even) $O(1)$ instanton, both terms in the quantisation condition must be integer classes by themselves. I.e. $F_E$ must be an integer class and $E$ must be spin.\footnote{For more general instantons, in particular for $U(1)$ instantons, this conclusion may be modified.} 
Both conditions are altered in the presence of background $B$-field as described in \cite{Collinucci:2008sq,Blumenhagen:2008zz,Grimm:2011dj}. 
As for the D7-brane world-volume fluxes, we can represent $\Sigma_E$ as the difference of holomorphic curves on $E$. Since $\Sigma_E$ must be odd under the orientifold involution, we can write $\Sigma_E=\calc_E-\calc_E'$, with $\calc_E'=\sigma_*\calc_E$.   

The gaugings (\ref{RRg}) of the R-R fields  imply that $e^{-S_{\rm D3}}$ transforms as a charged object under the D7-brane $U(1)$s \cite{Blumenhagen:2006xt,Ibanez:2006da,Florea:2006si,Haack:2006cy}
\be\label{shiftD3}
e^{-S_{\rm D3}}\quad\rightarrow \quad e^{\ii q_a\lambda^a}e^{-S_{\rm D3}}
\ee
with quantised charge given by \cite{Grimm:2011dj}
\be\label{E3charges}
q_a=  \Sigma_E\cdot D_a - E\cdot \Sigma_a\,.
\ee
Here we have used $\Sigma_E\cdot D_a=-\Sigma_E\cdot D'_a$ and $E\cdot \Sigma_a=-E\cdot \Sigma'_a$. In passing, notice also that $\Sigma_E\cdot \hat D_\alpha=0$ for a D7-brane along an orientifold even divisor $\hat D_\alpha$ since $\Sigma_E$ is odd. 

It is important to appreciate that     the charges (\ref{E3charges}) receive a contribution from the 7-brane flux \emph{and} the instanton flux
and that $e^{-S_{\rm D3}}$ is charged only under the combinations of $U(1)$s that are St\"uckelberg massive. 
In particular, since $\Sigma_E$ is odd and $E$ is even, we can write $\Sigma_E\cdot D_a\equiv\frac12\Sigma_E\cdot (D_a-D'_{a})$ and $E\cdot \Sigma_a\equiv \frac12 E\cdot (\Sigma_a-\Sigma'_a)$, 
which makes manifest how  $\Sigma_E\cdot D_a$ and $E\cdot \Sigma_a$ in  (\ref{E3charges}) are due to the geometric and flux-induced St\"uckelberg gauging (\ref{RRC2}) and (\ref{RRC4}), respectively.

The charges (\ref{E3charges}) are crucial for determining the selection rules governing the couplings that appear in the four-dimensional effective theory. Adopting the terminology of \cite{Martucci:2015dxa}, the $U(1)$ gauge symmetries that are St\"uckelberg massive remain as {\em perturbative} selection rules  for the F-terms involving  charged matter fields. On the other hand,  supersymmetric D3-instantons  can produce non-perturbative F-term couplings that are exponentially suppressed by $e^{-S_{\rm D3}}$ and violate such  perturbative selection rules, possibly reducing them to discrete ones \cite{BerasaluceGonzalez:2011wy}. Indeed, the charge of $e^{-S_{\rm D3}}$ must be compensated  by the contribution to the same coupling of matter fields with opposite overall charge under the St\"uckelberg massive
$U(1)$s  \cite{Blumenhagen:2006xt,Ibanez:2006da,Florea:2006si}. 
Let us review the microscopic origin of this effect.

The contribution of a D3-brane instanton to the four-dimensional effective theory is obtained by evaluating the corresponding partition function. In particular, the phase shift (\ref{shiftD3}) associated with the D3-brane classical action is compensated by the appearance of fermionic zero modes in the path-integral measure. Let us define the curves $\gamma_a=D_a\cap E$, $\gamma'_a=D_a'\cap E$ and $\hat\gamma_\alpha=\hat D_\alpha\cap E$. 
For instance $\gamma_a$ supports $n_a^+$ chiral fermions $\eta_{a}$ with charge $+1$  and  $n_a^-$  zero modes  $\tilde \eta_{ a}$ with charge $-1$ under the four-dimensional $ U(1)_{a}$ gauge group.\footnote{Since we are focusing on $O(1)$ instantons, there is no `zero-mode' contribution to the $U(1)$ D3-brane world-volume gauge symmetry, under which $\eta_{a}$ and $\tilde\eta_{a}$  would have charges $-1$ and $+1$ respectively, as is the case for $U(1)$ instantons.}
The exact number of such zero modes on the curve $\gamma_a$ is given by\footnote{Our conventions in (\ref{nzerom})  differ from the ones used e.g. in \cite{Grimm:2011dj} by a minus sign in order to make them compatible with our choice of sign for the R-R fields $C$ in (\ref{SD3conventions}).}
\be\label{nzerom}
n_a^+= {\rm dim}\, H^0 (\gamma_a, L_E\otimes L^{-1}_a  |_{\gamma_a} \otimes K^{1/2}_{\gamma_a}), \qquad n_a^-= {\rm dim}\, H^1 (\gamma_a,L_E\otimes L^{-1}_a |_{\gamma_a} \otimes K^{1/2}_{\gamma_a}).
\ee
Here $L_E$  and $L_a$ are the line bundles with curvature $\frac{1}{2\pi} F_E$ and $\frac{1}{2\pi} F_a$, respectively, and $K^{1/2}_{\gamma_a}$ defines the spin structure on the curve $\gamma_a$. 
Similarly there exist  $n^{\prime +}_{a}$ and  $n^{\prime -}_{a}$ zero-modes  $\eta'_{a}$ and $\tilde \eta_{a}'$ on $\gamma'_{a}$, and $\hat n^+_\alpha$ and  $\hat n^-_\alpha$  zero-modes $\eta_{\alpha}$ and $\tilde \eta_{\alpha}$ on $\hat\gamma_{\alpha}$, counted by corresponding cohomology groups.
The chiral index of such  charged zero modes of a fluxed D3-brane instanton is therefore given by \cite{Grimm:2011dj}
\be \label{chiralindexa}
n^+_a-n^-_a=n^{\prime +}_{a}-n^{\prime -}_{a}=\frac1{2\pi}\int_{\gamma_a}(F_E-F_a)\equiv q_a
\ee
in agreement with (\ref{E3charges}).
Furthermore $\hat n^+_\alpha-\hat n^-_\alpha=\frac1{2\pi}\int_{\hat \gamma_{ \alpha}}(F_{E}- \hat F_{ \alpha})=\Sigma_E\cdot \hat D_\alpha-\hat \Sigma_\alpha\cdot E=0$, since $\Sigma_E$ and $\hat \Sigma_{ \alpha}$ are odd, while $\hat D_\alpha$ and $E$ are even. 

Because of the orientifold projection only, say,  $\eta_{a},\tilde \eta_{ a},  \eta_{\alpha},\tilde\eta_{\alpha}$, contribute to the D3-instanton path-integral measure
\be
\prod_a \d\eta_a\d\tilde\eta_a\prod_\alpha \d\eta_\alpha\d\tilde\eta_\alpha,
\ee
which, under the $U(1)_a$ gauge symmetries, transforms with a phase $e^{-\ii q_a\lambda^a}$. 
Of course, in absence of D3-brane interactions that can `soak-up' such zero modes,  the instantonic D3-brane partition function, although formally gauge invariant, identically vanishes.

On the other hand, if two curves, say, $\gamma_a$ and $\gamma_b$ intersect at points $p_{ab}=\gamma_a\cap \gamma_b$, 
the chiral fermions couple to bosonic charged matter zero-modes $\Phi_{ab}$ localised on the matter curves $D_a\cap D_b$ of charge $(-1,1)$  under  $U(1)_a\times U(1)_b$.
The instanton effective action can then include gauge invariant interaction terms of the schematic form
\be\label{intterms}
\eta_{a} \, \Phi_{ab} \, \tilde \eta_{b} \,.
\ee

Imagine for instance  two D7-branes such that the instanton has charges $q_a=-q_b=n$, and suppose furthermore that this charge counts the exact number of charged fermionic zero modes $\eta_a$ and $\tilde\eta_b$, so that they do not enter in any vectorlike pair. Hence, 
 the  term (\ref{intterms}) can soak up the $n$ fermionic zero modes of type $\eta_{a}$ and $\tilde \eta_{b}$, producing an F-term proportional to 
 \be \label{Phi-coup1}
 (\Phi_{ab})^n\, e^{-S_{\rm D3}} .
 \ee
The same argument can be repeated for interactions involving other chiral fermions and four-dimensional chiral fields.  See also the review \cite{Blumenhagen:2009qh} for more information on such instanton induced couplings.
Note that the appearance of a suitable coupling of the type (\ref{Phi-coup1}) depends on the exact number $n^+_a, n^-_a$ of charged instanton zero modes rather than merely their chiral index.

Finally, notice that even in presence of 7-brane gauge fluxes, there could be a cancellation of the D3-instanton charges $q_a$ so that no insertion of chiral fields is needed. This happens if 
\be
E\cdot \Sigma_a=\Sigma_E\cdot D_a.
\ee
According to the above discussion, this cancellation can happen only if the $U(1)_a$ gauge symmetry is massive by  both the geometric and flux-induced   St\"uckelberg mechanisms. This effect can have important consequences e.g. for the role of the instanton in moduli stabilisation \cite{Grimm:2011dj} as the instanton flux can help in overcoming the challenge pointed out in \cite{Blumenhagen:2007sm}. One of our goals is to understand this phenomenon in F-theory.

\section{A non-perturbative Yukawa coupling via the interplay of instanton and gauge flux in a Type IIB setting} \label{Appendix-IIBExample}

In this appendix we come back to the non-perturbative generation of the ${\bf 10 \, 10 \, 5}$ Yukawa coupling in $SU(5)$ GUT models discussed at the end of section \ref{sec:G4trivial}, but focus on the analogous D3-instanton configuration in the weakly coupled Type IIB limit. The fact that in Type II orientifolds instantons can generate this coupling is well-known \cite{Blumenhagen:2007zk}. Our goal here is to exemplify the interplay between the gauge and the instanton flux in a concrete example. This way we illustrate that the difficulty of finding an explicit supersymmetric 3-form flux $T_3$ in M-theory translates into the challenge of constructing a line bundle on the instanton with the correct amount not only of chiral, but also of non-chiral charged zero modes.

Concretely, consider the $SU(5)$ Tate model presented in \cite{Blumenhagen:2009up} in its Sen limit. At weak coupling, this model reduces to a Type IIB orientifold compactification on a Calabi-Yau 3-fold $X$ with three divisor classes $H$, $U$, $V$ together with an orientifold involution exchanging $U$ and $V$.
This geometry has also been used in section 6.5 of \cite{Martucci:2015dxa}, to which we refer for more details. The non-trivial intersection numbers of the divisor classes on $X$ are given as 
$H^2 U = 2$, $H U^2 = -2$, $U^3= 2$, $H U V =1$, $U^2 V = -1$ (plus the ones obtained by exchanging $U$ and $V$). 

A stack of 5 coincident 7-branes on $U$ together with the orientifold image stack on $V$ give rise to a gauge group $U(5)$. The 7-brane tadpole is canceled by an invariant 7-brane along a Whitney-type divisor $\Sigma$ in class $8 H + 3 (U+V)$.
Charged matter in the representation ${\bf 10}_2$ resides on the curve ${C}_{\bf 10} = U \cap V$ and an extra representation ${\bf 5}_1$ is localised on ${C}_{\bf 5} = U \cap \Sigma$.
It would be easiest to focus on a situation with vanishing gauge fluxes. However, recall that the Freed-Witten quantisation condition enforces a gauge flux on a 7-brane along a non-spin divisor $Q$ in such a way that
$c_1(L_Q) + \frac{1}{2} c_1(K_Q) $ is an integer class. Here $L_Q$ denotes the line bundle on $Q$ with $c_1({L_Q}) = \frac{1}{2\pi} F_Q$, and furthermore  
$c_1(K_Q) = Q$ for a divisor $Q$ on a Calabi-Yau. Since the divisor in class $U$ is a del Pezzo surface and hence non-spin, this means that $F_U = \frac{1}{2} U + \tilde F_U$ for an integer class $\tilde F_U$, and similarly for $F_V$.\footnote{Since $U+V$ is an even class, no gauge flux is required by the quantisation condition on $\Sigma$.}

Consider now a D3-instanton on an invariant divisor $E$ in class $E = a (U +V) + b H$ for $a$, $b$ to be determined.
The quantisation condition for instanton flux requires that $F_E + \frac{1}{2}(a U + a V + b H)|_E$ be integer. Since $F_E$ is orientifold odd, the quantisation condition is satisfied
 if $a, b \in 2 \mathbb Z$ and $F_E$ integer.
According to  the discussion around (\ref{chiralindexa}) a necessary condition for this instanton to contribute to a coupling of the form ${\cal O}^n = [{\bf 10}_2 \, {\bf 10}_2 \, {\bf 5}_1]^n$ is that there exists instanton flux $\frac{1}{2\pi} \, F_E \in H^{1,1}_-(E)$ such that
\be \label{necconda}
\frac{1}{2\pi} \int_{\gamma_{U}} (F_E -  F_U) = - n, \qquad \quad \gamma_U = E \cap U.
\ee 
This is the condition for a  chiral excess of charged zero modes $\tilde \eta_U^a$, $a=1, \ldots, n$, each transforming as a ${\bf \bar 5}_{-1}$ of $U(5)$ on the intersection locus  $\gamma_U$. The charge of these chiral zero modes cancels the charge $5 n$ of the operator ${\cal O}^n$. This, however, is merely a necessary condition for the operator to be actually induced the D3-instanton. In addition there must exist one zero mode $\eta_\Sigma$ in the sector $E - \Sigma$ on $\gamma_\Sigma = \Sigma \cap E$, which is mapped to its vector-like partner $\tilde \eta_\Sigma$ in the sector $\Sigma-E$. Furthermore, all extra vectorlike pairs of zero modes, if present, must be absorbable by instanton couplings. A sufficient condition is therefore to require that no such zero modes be present. This translates into the conditions
\be
\begin{aligned}
& h^0(\gamma_U, L_E \otimes L_U^{-1}|_{\gamma_U} \otimes K^{1/2}_{\gamma_U}) = 0\,, \qquad & h^1(\gamma_U, L_E \otimes  L_U^{-1}|_{\gamma_U} \otimes K^{1/2}_{\gamma_U}) =n\,, \\
& h^0(\gamma_\Sigma, L_E \otimes \otimes L_\Sigma^{-1}|_{\gamma_\Sigma} \otimes  K^{1/2}_{\gamma_\Sigma}) = 1\,, \qquad & h^1(\gamma_\Sigma, L_E \otimes L_\Sigma^{-1}|_{\gamma_\Sigma}   \otimes K^{1/2}_{\gamma_\Sigma}) =1\, .
\end{aligned}
\ee

To come back to the necessary condition (\ref{necconda}), on the concrete 3-fold $X$ an ansatz for the instanton flux is $\frac{1}{2 \pi}F_E =c \, (U-V)|_E$ for $c \in \mathbb Z$. 
We then make the general ansatz $F_U = ((x+\frac12) U + y V + z H)|_U$ with $x,y,z \in \mathbb Z$ and $F_\Sigma = \frac{1}{2} (2 \tilde p +1) (U-V)|_\Sigma$ with $\tilde p \in \mathbb Z$.
The D5-tadpole cancellation condition
\be
F_U \wedge U + F_V \wedge V + F_\Sigma \wedge \Sigma = 0   \in H^4(X)
\ee
relates these coefficients as $x-y = 5 \tilde p +2$.
The intersection numbers together with this constraint then imply that
\be
\frac{1}{2\pi} \int_{\gamma_U} (F_U - F_E) = 
 \frac{5}{2} a - 3 b - 3 a \, c + 3 b \, c + 5 a \, \tilde p - 10 b \, \tilde p - a \, y - b \, y - 
 a \, z + 2 \, b \, z,
\ee
which is integer because $a \in 2 \mathbb Z$. It is easy to find solutions with $n=1$. More work is required, however, to find solutions which also satisfy the conditions on the vector-like zero modes (including the deformation modes if the instanton divisor is non-rigid). It is expected that the uplift of such solutions then describes a consistent M5-flux $T_3$ which satisfies the Bianchi identity (\ref{BIref}).
Note in particular that $F_U$ corresponds to the massive $U(1)$ gauge flux. Consistency requires that the pullback of its F-theory uplift to the M5-divisor $D$ is indeed cohomologically trivial on $D$.  In section \ref{sec_thirdcase} this is precisely what we show in the stable degeneration limit of a prototypical massive $U(1)$ model.

\newpage

\newpage
\bibliography{references}  
\bibliographystyle{custom1}

\end{document}